\documentclass[iop]{emulateapj}
\usepackage{txfonts}
\usepackage{graphicx}
\usepackage{natbib}

\shorttitle{Extended Stellar Components of Galaxies \& the Nature of Dark Matter}
\shortauthors{Power \& Robotham}

\begin{document}


\title{The extended stellar component of galaxies \& the nature of dark matter}


\author{C. Power\altaffilmark{1} \& A. S. G. Robotham}
\affil{International Centre for Radio Astronomy Research,
  University of Western Australia, 35 Stirling Highway, Crawley, 
  WA 6009, Australia}


\altaffiltext{1}{E-mail:~\texttt{chris.power@icrar.org}}


\def\lesssim{\mathrel{\hbox{\rlap{\hbox{\lower4pt\hbox{$\sim$}}}\hbox{$<$}}}}
\def\gtrsim{\mathrel{\hbox{\rlap{\hbox{\lower4pt\hbox{$\sim$}}}\hbox{$>$}}}}


\begin{abstract}
  Deep observations of galaxies reveal faint extended stellar components
  (hereafter ESCs) of streams, shells, and halos. These are a natural
  prediction of hierarchical galaxy formation, as accreted satellite galaxies
  are tidally disrupted by their host. We investigate whether or not global
  properties of the ESC could be used to test of dark matter, reasoning that
  they should be sensitive to the abundance of low-mass satellites, and
  therefore the underlying dark matter model. Using cosmological simulations
  of galaxt formation in the favoured Cold Dark Matter (CDM) and Warm Dark
  Matter (WDM) models ($m_{\rm WDM}=0.5,1,2 {\rm keV}/c^2$), which suppress the
  abundance of low-mass satellites, we find that the kinematics and orbital
  structure of the ESC is consistent across models. However, we find striking
  differences in its spatial structure, as anticipated -- a factor of
  $\sim\,10$ drop in spherically averaged mass density between $\sim\,10\%$ and
  $\sim\,75\%$ of the virial radius in the more extreme WDM runs
  ($m_{\rm WDM}$=0.5, 1 ${\rm keV}/c^2$) relative to the CDM run. These
  differences are consistent with the mass assembly histories of the
  different components, and are present across redshifts. However, even the
  least discrepant of the WDM models is incompatible with current observational
  limits on $m_{\rm WDM}$. Importantly, the differences we observe when varying
  the underlying dark matter are comparable to the galaxy-to-galaxy variation
  we expect within a fixed dark matter model. This suggests that it will be
  challenging to place limits on dark matter using only the unresolved spatial
  structure of the the ESC.
\end{abstract}


\section{Introduction}
\label{sec:intro}

Arguably the defining prediction of the canonical Cold Dark Matter (CDM) model
of cosmological structure formation is that dark matter haloes should contain an
abundance of low-mass substructure haloes
\citep[hereafter subhalos; e.g.][]{klypin.etal.1999b,moore.etal.1999,reed.etal.2005,springel.etal.2008,ishiyama.etal.2013},
remnants of the merging hierarchy through which their hosts assembled. Cosmological
$N$-body simulations have revealed that the number density of these subhalos increases
with decreasing subhalo mass scale \citep[e.g.][]{reed.etal.2005}, approximately as
$n_{\rm sub} \propto M_{\rm sub}^{-\alpha}$, with $\alpha \simeq 1.9$
\citep[e.g.][]{gao.etal.2004,diemand.etal.2007,springel.etal.2008,garrison-kimmel.etal.2014},
and this applies equally in the dark matter host halos of massive
galaxy clusters of $10^{15} {\rm M}_{\odot}$ down to the hosts of galaxies like the
Milky Way ($\sim 10^{12} {\rm M}_{\odot}$) and dwarfs
\citep[e.g.][]{angulo.etal.2009,xie.gao.2015,rodriguez-puebla.etal.2016}.

Alternatives to the CDM model, such as Warm Dark Matter (WDM), suppress the abundance
of low-mass dark matter halos, and consequently the abundance of subhalos,
\citep[e.g.][]{smith.markovic.2011,schneider.2013,benson.2013,pacucci.2013}, but
distinguishing between these alternatives and CDM in a robust fashion has proven
challenging \citep[e.g.][]{knebe.etal.2008}. The mass scale at which differences
between plausible dark matter models is most
likely to be evident -- at or below the scale of the satellites of the Milky Way 
\citep[e.g.][]{anderhalden.etal.2013,schneider.etal.2014} -- is also the mass scale
at or below which galaxy formation is inefficient and apparently stochastic
\citep[e.g.][]{boylan-kolchin.etal.2011,power.2014,garrison-kimmel.etal.2016}. The latest generation of
cosmological hydrodynamical galaxy formation simulations in the CDM model can produce
satellite populations that are broadly consistent with observations
(e.g. \citealt{brooks.zolotov.2014}, \citealt{sawala.etal.2016a}, \citealt{sawala.etal.2016b},
\citealt{dutton.etal.2016}, \citealt{wetzel.etal.2016}, \citealt{zhu.etal.2016}).
However, it is noteworthy that the internal properties of galaxies and their satellites in plausible
alternatives, such as WDM or Self-Interacting Dark Matter (SIDM), can also provide similarly good
consistency with observations
(e.g. \citealt{zavala.etal.2013}, \citealt{herpich.etal.2014},
\citealt{colin.etal.2015}, \citealt{elbert.etal.2015}, \citealt{fry.etal.2015},
\citealt{governato.etal.2015}).

These results suggest that, provided a galaxy can form, the properties of the galaxy that
we observe are likely to be shaped by the physics of galaxy formation rather than the physics
of dark matter \citep[e.g.][]{herpich.etal.2014}. Arguably this is not so surprising, when one
considers the relative dominance of baryonic material for the central gravitational potential in all but
the most dark matter dominated galaxies, and the relatively short timescales (e.g. crossing times,
cooling times) in these regions. What about the outskirts of galaxies, where timescales are long
and the influence of dark matter dominates that of baryons? Could these regions offer a test of
the underlying dark matter model?

\medskip

Deep observations of the Milky Way and external galaxies reveal that
they are embedded in diffuse, extended stellar structures of shells, streams, and
halos \citep[][]{freeman.bland-hawthorn.2002,helmi.2008}. These structures are strikingly evident
around our nearest massive galactic neighbour, M31 -- see, for example, results from
the {\small PAndAS} \citep[e.g.][]{mcconnachie.etal.2009,mackey.etal.2010} and {\small SPLASH} surveys
\citep[e.g.][]{gilbert.etal.2012} -- and they are also apparent in deep imaging of more distant galaxies
\citep[e.g.][]{martinez-delgado.etal.2010,radburn-smith.etal.2011,monachesi.etal.2013,trujillo.etal.2013,crnojevic.etal.2015}.
Stellar halos are predicted to arise naturally
in hierarchical galaxy formation models, built up, at least partially, by the merger
and accretion events that drive galaxy assemly \citep[e.g.][]{searle.zinn,helmi.white.1999,bullock.johnston.2005,abadi.etal.2006,johnston.etal.2008,zolotov.etal.2009,cooper.etal.2010,font.etal.2011,scannapieco.etal.2011,mccarthy.etal.2012,cooper.etal.2013}; partially, because cosmological galaxy formation simulations have
demonstrated that such halos likely have a dual origin, comprising of a component that
formed in-situ \citep[cf.][]{abadi.etal.2006,zolotov.etal.2009,font.etal.2011}, in
addition to the accreted component that formed via tidal disruption of satellite galaxies
\citep[e.g.][]{bullock.johnston.2005} and star clusters \citep[e.g.][]{boley.etal.2009}.
The in-situ component is believed to originate in dynamical heating of the stellar disc
at early times \citep[cf.][]{mccarthy.etal.2012}, and so it dominates at smaller
galacto-centric radius; in contrast, the outer halo is dominated by the accreted
component (e.g. \citealt{zolotov.etal.2009}, \citealt{scannapieco.etal.2011},
\citealt{amorisco.2015}, \citealt{rodriguez-gomez.etal.2016}).

In this paper, we investigate how the underlying dark matter model might influence this accreted
component by exploring the spatial and kinematic
structure of stellar halos of simulated galaxies, to which we refer hereafter as extended
stellar components (ESCs), and focus on those that form in
cosmological galaxy formation simulations in CDM and WDM models. We reason that there
should be an imprint on properties of the accreted component; because this component
is built up through mergers and accretions of lower mass satellite galaxies, it
follows that a dark matter model that suppresses the abundance of low-mass subhalos, and
consequently low-mass satellite galaxies, is likely to result in less massive -- and
lower luminosity -- ESCs at large galacto-centric radii, or equally, more
centrally concentrated ESCs in WDM models compared to the CDM counterpart.

We do not expect the dynamics of subhalos to differ significantly between CDM and WDM
models \citep[e.g.][]{knebe.etal.2008}. The orbital distribution will be preferentially
radial \citep{benson.2005}, while the orbits of lower mass subhalos and their satellites
will
require many pericentric passages to decay \citep{tormen.etal.1998}; this implies that
tidally stripped stars from a low-mass satellite can be spread over large galacto-centric
distances, and the greater the number of low-mass satellite undergoing such tidal
stripping, the greater the radial extent of the resulting ESC. We make the reasonable
assumption that the physical processes that govern galaxy formation (e.g. cooling, star
formation, feedback) should not depend on the underlying dark matter, consistent with
previous studies \citep[e.g.][]{herpich.etal.2014,governato.etal.2015}. This means that
differences in the properties of the accreted component of the ESC should
reflect differences in the abundance of satellites, which depends on the subhalo
abundance. These differences could be accessible to future surveys that target the
diffuse, low surface brightness environs of galaxies, extending the work of deep imaging
studies such as {\small PAndAS} \citep[cf.][]{mcconnachie.etal.2009} and {\small SPLASH}
\citep[e.g.][]{gilbert.etal.2012} and the the ``Dragonfly Telephoto Array''
\citep[cf.][]{dragonfly}, and so potentially represents an observational test of dark matter.

\medskip

In the remainder of this paper, we present our exploration of this idea, using
cosmological zoom simulations of a set of six Milky Way mass system
($M_{200} \simeq 2 \times 10^12 h^{-1} {\rm M}_{\odot}$). In all six cases, we
follow their formation and evolution in a fiducial CDM model, and for one system
we carry out a further set of simulations in three WDM models, with equivalent
particle masses of $m_{\rm WDM}$=0.5, 1 and 2 ${\rm keV}/c^2$; we motivate these choices
below. All simulations assume
the same prescriptions for cooling, star formation and supernova feedback, and we carry
out a set of simulations to assess the sensitivity of our predictions to the assumed
galaxy formation parameters. Details of the simulations are presented in
\S\ref{sec:sims}. Results of our analysis are given in \S\ref{sec:results}; here we
quantify the $z$=0 spatial and kinematic structure of the ESC; verify that
these results are reasonable in the context of the mass assembly histories of the
galaxies and the orbital structure of the ESC; and estimate whether or not the
differences could provide a useful test of dark matter, allowing for the system-to-system
variation we might expect within the CDM model. Finally, in \S\ref{sec:summary},
we summarise our results.

\section{Simulations}
\label{sec:sims}

\paragraph*{Parent Simulation} The parent $\Lambda$CDM $N$-body simulation is a
$256^3$ particle $L_{\rm box}=50 h^{-1} \rm Mpc$ cube. Following \citet{komatsu.etal.2011},
we assume total matter, baryon, and dark energy density parameters of $\Omega_{\rm m}=0.275$,
$\Omega_{\rm b}=0.0458$, and $\Omega_{\Lambda}=0.725$, and dimensionless Hubble parameter of
$h=0.702$, and we compute the CDM power spectrum using {\texttt{CAMB}} \citep[cf.][]{lewis.etal.2000},
with a power spectrum normalisation of $\sigma_8=0.816$ and primordial spectral index
$n_{\rm s}=0.968$. This results in a particle mass of $m_p \simeq 5.6 \times 10^8 h^{-1} \rm M_{\odot}$.
The simulation is run with {\small GADGET2} \citep{springel.2005} with a constant comoving
gravitational softening $\epsilon=5 h^{-1} \rm kpc$, and candidate haloes for resumulation
are identified in the friends-of-friends (FOF) group catalogue at $z$=0, where we use a
linking length of $b$=0.2$\bar{d}$. Note that we post-process each FOF group to compute its centre of
density $\vec{r}_{\rm cen}$
\citep[using the iterative ``shrinking spheres'' method; cf.][]{power.etal.2003} and obtain
virial masses, which we define as
\begin{equation}
  M_{200} = \frac{4\pi}{3}\times 200 \times \rho_{crit,0} R_{200}^3;
\end{equation}
here $\rho_{\rm crit,0}=2.7755\times10^{11}{\rm M}_{\odot}h^{2}{\rm Mpc}^{-3}$ is the critical density
of the Universe at $z$=0. We select Milky Way mass haloes as those having virial masses of
$M_{200} \sim 2\times 10^{12} h^{-1} \rm M_{\odot}$, which ensures that they are resolved at $z$=0 with
$\sim 3,500$ particles; this is sufficient to define accurately the Lagrangian region at high
redshifts from which the halo collapses.

\begin{table*}
\begin{center}
  \caption{\textbf{Galaxy Properties at $z$=0.} For each of the galaxies that form in our runs with fiducial galaxy formation parameters, we give (1) the virial mass, ${M}_{200}$, as defined in the text; (2,3) $f_{\rm g}$ and $f_{\rm s}$, the fraction of ${\rm M}_{200}$ in gas and stars, respectively; (3,4,5) $N_{\rm d}$, $N_{\rm g}$, $N_{\rm s}$, the number of dark matter, gas, and star particles within ${\rm R}_{200}$; and (6,7,8) $\sigma_{\rm 3D}^{\rm d}$, $\sigma_{\rm 3D}^{\rm g}$, and $\sigma_{\rm 3D}^{\rm s}$, the 3D velocity dispersions of all the dark matter, gas, and star particles within ${\rm R}_{200}$.}
\vspace*{0.3 cm}

\begin{tabular}{lcccccccccc}\hline
  & $M_{200}$   & $f_{\rm g}$& $f_{\rm s}$ & $N_{\rm d}$& $N_{\rm g}$& $N_{\rm s}$ & $\sigma_{\rm 3D}^{\rm d}$ & $\sigma_{\rm 3D}^{\rm g}$ & $\sigma_{\rm 3D}^{\rm s}$\\
  & [$h^{-1} {\rm M}_{\odot}$] &&&&&[km/s]&[km/s]&[km/s]\\
  \hline
MW01 & $2.95 \times 10^{12}$&  0.095 & 0.077 & 518,571 & 298,102 & 242,495 & 303 & 149 & 410\\
MW02  &                      &        &       &         &         &         &
&    &    \\
\hline
WDM0.5 & $2.19 \times 10^{12}$&  0.099 & 0.083 & 374,586 & 227,325 & 190,160 & 293 & 122 & 329\\
WDM1 & $2.28 \times 10^{12} $ &  0.098 & 0.079 & 392,963 & 233,625 & 190,266 & 302 & 156 & 327 \\
WDM2 & $2.203 \times 10^{12}$ &  0.095 & 0.077 & 380,905 & 217,819 & 176,575 & 298 & 137 & 324 \\
CDM & $2.26 \times 10^{12}$   &  0.086 & 0.096 & 385,618 & 204,528 & 227,176 & 319 & 134 & 369 \\
\hline
MW03 & $2.24 \times 10^{12} $ &  0.115 & 0.079 & 392,963 & 233,625 & 190,266 & 302 & 156 & 327 \\

MW04 & $2.45 \times 10^{12}$ &  0.097 & 0.08 & 370,748 & 217,678 & 179,171 & 296 & 147 & 383 \\

MW05 & $2.33 \times 10^{12}$ &  0.108 & 0.077 & 383,492 & 253,462 & 167,170 & 298 & 127 & 368 \\

MW06 & $2.54 \times 10^{12}$ &  0.119 & 0.048 & 392,423 & 279,630 & 111,835 & 265 & 151 & 292 \\

\hline
\end{tabular}
\label{tab:assembly_histories}
\end{center}
\end{table*}

\bigskip

\paragraph*{Galaxy Resimulations} We resimulate a suite of 6 Milky Way mass halos (MW01-MW06; see
Table~\ref{tab:assembly_histories} for details) with both dark matter and gas. They were chosen to
reside in low-density (void) regions, identified using the {\small V-web} algorithm of
\citet{hoffman.etal.2012}, which is a kinematic classification of the cosmic web based
on diagonalisation of the local velocity shear tensor. Our approach to generating initial
conditions is presented in detail in \citet{power.etal.2014}, but we summarise the key
steps as follows.

\begin{enumerate}
\item We select all particles within radius $5 R_{200}$ at $z$=0 centred on
  $\vec{r}_{\rm cen}$ of the halo of interest within the parent simulation, and use
  these to define the Lagrangian volume encompassing the desired high resolution region in the
  resimulation initial conditions at $z=\infty$.
\item We populate this high resolution region with both dark matter and gas
  particles whose relative mass densities are set by the baryon and dark matter density
  pararmeters $\Omega_{\rm b}=0.0458$ and $\Omega_{\rm DM}=0.2292$, and whose number densities are
  set by the desired maximum mass resolution of the resimulation. This high resolution region is
  then embedded within regions of coarser mass resolution consisting of collisionless tidal
  particles, where the average particle mass increases with increasing distance from the
  centre of the high resolution patch.
\item We choose a starting redshift $z_{\rm start}$, in this case $z_{\rm start}=99$, and
  impose the appropriate set of density perturbations on the composite particle
  distribution.

  The first set are applied to all particles and are the original perturbations
  used in the parent simulation, with minimum and maximum wavenumbers, $k_{\rm min}=2\pi/L_{\rm box}$ and
  $k_{\rm max}=\pi\,N_{\rm parent}/L_{\rm box}$; here $L_{\rm box}$ is the length of the parent cube
  and $N_{\rm parent}$ is the number of particles on a side within this cube.

  The second set are applied to the particles in the high resolution region and are a new set of
  perturbations that were not present in the initial conditions; note that we sample from
  the appropiate baryon and dark matter power spectra for the gas and dark matter particles separately.
  Here the minimum and maximum wavenumbers are $k_{\rm min}=2\pi/L_{\rm hires}$ and
  $k_{\rm max}=\pi\,N_{\rm hires}/L_{\rm hires}$, where $L_{\rm hires}$ is the length of the cube
  encompassing the high resolution patch and $N_{\rm hires}$ is the number of particles
  on a side within this cube.
  
\item Finally, we use the Zel'dovich approximation to generate particles'
  initial displacements and velocities \citep[cf.][]{zeldovich.1970,efstathiou.etal.1985}.
\end{enumerate}
\medskip
In the case of MW02, we run a series of simulations in which we vary galaxy formation
parameters and numerical resolution (see below), and, crucially for this study, the
underlying dark matter model.
Following \citet{bode.etal.2001}, we obtain the initial power spectra
for our WDM models by filtering the CDM power spectrum with an additional transfer
function of the form
\begin{equation}
  \label{eq:transfer}
  T^{\rm WDM}(k) = \left(\frac{P^{\rm WDM}(k)}{P^{\rm CDM}(k)}\right)^{1/2}=\left[1+(\alpha\,k)^{2\nu}\right]^{-5/\nu}
\end{equation}
where $\alpha$ is a function of the WDM particle mass \citep[see equation A9 of][]{bode.etal.2001}, $k$ is the wave-number and $\nu$=1.2 is a numerical constant. We do
not include an additional velocity to mimic the effects of free-streaming in the
early Universe. Arguably this omission is likely to be unimportant for the
WDM particle masses we consider \citep[e.g.][]{colin.etal.2008,angulo.etal.2013}, but
we note also that modelling this effect correctly in a $N$-body simulation is
difficult -- it can lead to an unphysical excess of small-scale power in the initial
conditions if the simulation is started too early (see Figure 1 of
\citealt{colin.etal.2008} for a nice illustration of this problem) -- so for
clarity we ignore this effect \citep[see also discussion in][]{power.2013}.

In running our suite of resimulations, we use a version of {\small GADGET3} that has been
extended to model various galaxy formation processes, which we describe briefly in the following
subsection. Our version of {\small GADGET3} solves the equations of hydrodynamics using 
SPHS, a form of SPH that includes a higher order dissipation switch
\citep[SPHS; cf.][]{read.hayfield.2012,hobbs.etal.2013,power.etal.2014,sembolini.etal.2016}
with a Wendland $C^4$ kernel with 200 neighbours \citep[cf.][]{dehnen.aly.2012}.
Gravitational force softenings are set using the \citet{power.etal.2003} criterion,
$\epsilon_{\rm opt}=4\,R_{200}/\sqrt{N_{200}}$, where $N_{200}$ is the number of particles
within $R_{200}$ (cf. Table~\ref{tab:assembly_histories}).

\bigskip
\paragraph*{Galaxy Formation Prescription}
We follow the prescriptions for cooling, star formation, and supernova feedback
set out in \citet{hobbs.etal.2013}, which we now describe briefly.
Gas cools radiatively at temperatures above $10^4$ following \citet{katz.etal.1996},
assuming primordial abundances, and between $10^4$ and a floor of
$T_{\rm floor}$=$10^2$K following the prescription of \citet{mashchenko.etal.2008}.
These simulations do not include chemical evolution, which will influence cooling
rates as the abundance of heavier elements increases, especially in higher density
regions, but we are interested principally in relative differences and adopt the
same cooling rates in all of the resimulations.

Gas is prevented from cooling to the point at which the Jeans mass for gravitational
collapse becomes unresolved. We quantify mass resolution as
$M_{\rm res} = N_{\rm res}\,m_{\rm gas}$, where $m_{\rm gas}$ is the mass of a gas
particle and $N_{\rm res}$ is the number of gas particles that correspond to a single
resolution element; this is set to $N_{\rm res}=128$, which is reasonable for our
choice of smoothing kernel \citep[for further discussion, see][]{hobbs.etal.2013}.
This ensures that the Jeans mass is always resolved within our simulations, and we
write the Jeans density as
\begin{equation}
  \label{eq:jeans_density}
  \rho_J=\left(\frac{\pi\,k\,T}{\mu\,m_p\,G}\right)^3M_{\rm res}^{-2}
\end{equation}
where $k$ is Boltzmann's constant, $G$ is the Gravitational constant, $T$ is gas
temperature, $m_p$ is the proton mass, and $M_{\rm res}$ is the mass resolution, as
defined above. This manifests as a polytropic equation of state $P=A(s)\rho^{4/3}$,
where $s$ is the entropy; gas is prevented from collapsing to densities higher than
given by Eq~\ref{eq:jeans_density}, and any gas that lies on the polytrope forms
stars above a fixed density threshold with an efficiency of $\eta$=0.1, in accord
with observations of giant molecular clouds\citep[e.g.][]{lada.lada.2003}. The
star formation rate follows the \citet{schmidt.1959}
and \cite{kennicutt.1998} relation, $\rho_{\rm SFR}\propto \rho_{\rm gas}^{3/2}$, which
we implement by employing the dynamical time as the relevant star formation
timescale, i.e.
\begin{equation}
  \label{eq:dsfdt}
  \frac{d\rho_{\ast}}{dt}=\eta\frac{\rho_{\rm gas}}{t_{\rm dyn}}.
\end{equation}
We include feedback from supernovae (SNe) by injecting thermal energy from active
star particles into nearby gas particles. Finite resolution implies that each star
particle represents a single stellar population, which we assume to have a
\citet{salpeter.1955} initial mass function, and they become active once the age of
the star particle exceeds the mean main sequence age of stars more massive than 8
$\rm M_{\odot}$ and less massive than the maximum value of 100 $\rm M_{\odot}$. At
that point energy equivalent to $N_{\rm SNe}$ times the individual supernova energy
of $E_{\rm SN}=10^{51}$ ergs is injected as a delta function in time into a mass of
$M_{\rm res}$ gas particles, thereby ensuring that both star formation and feedback
is resolved.

\section{Results}
\label{sec:results}

\begin{figure*}
  \centerline{
    \plottwo{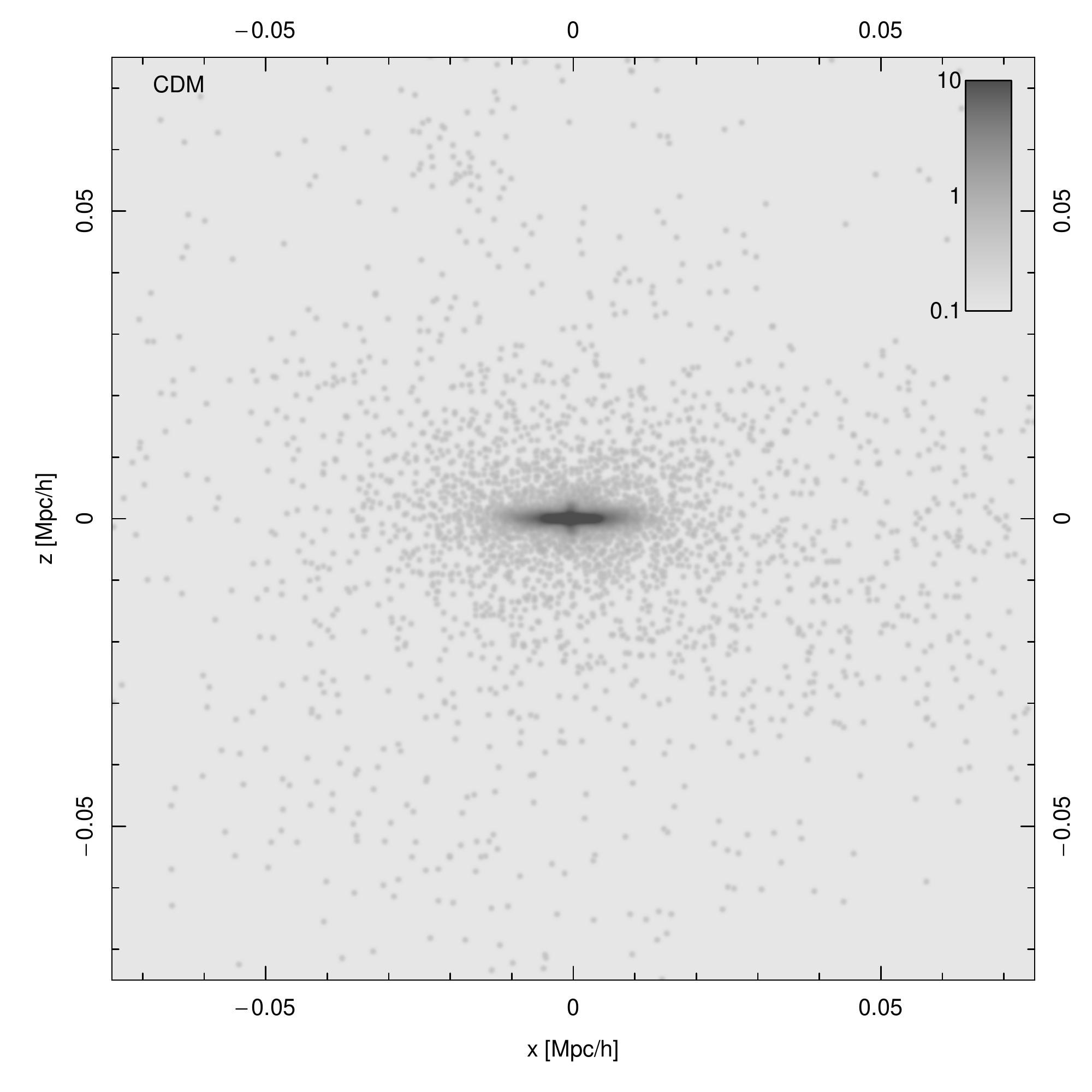}{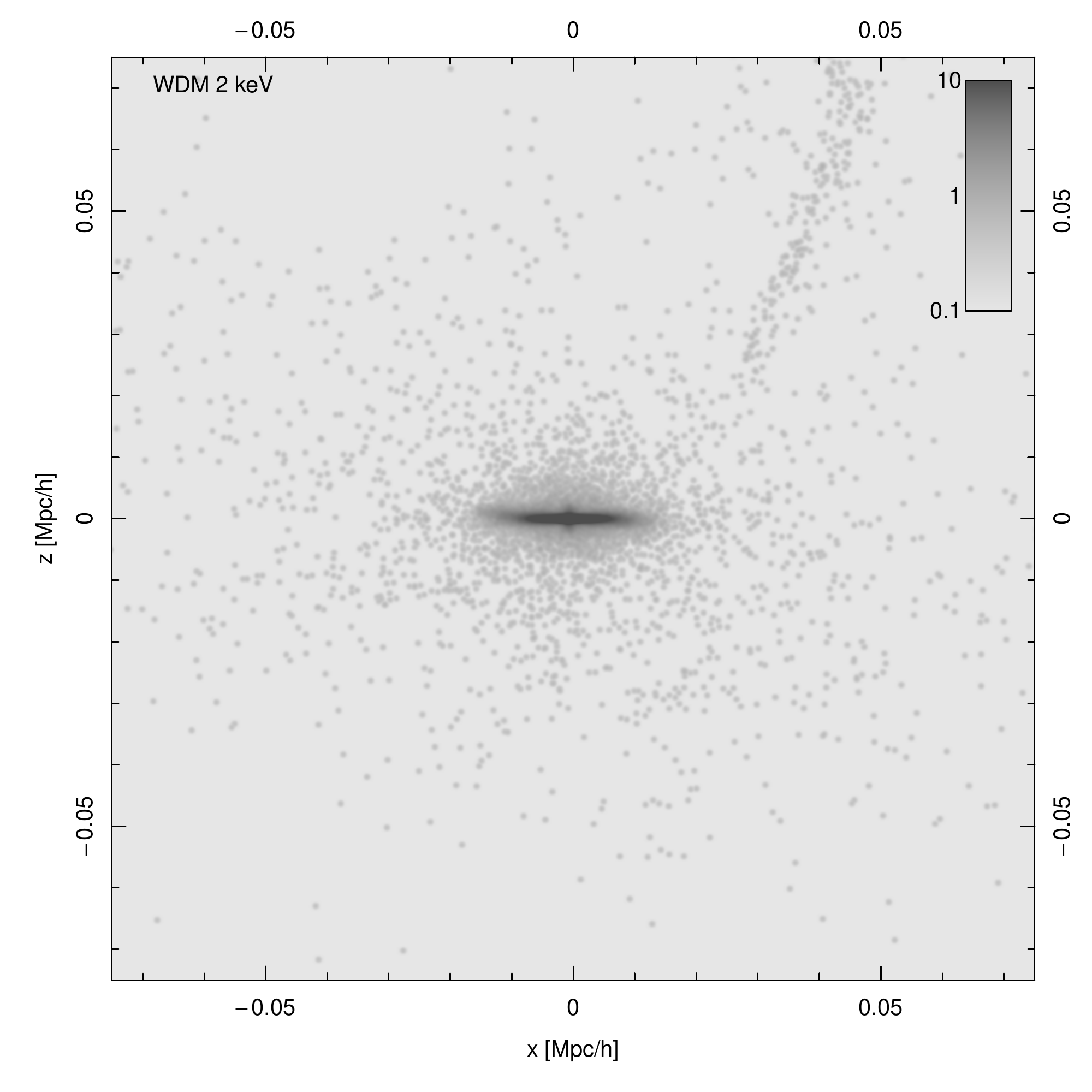}
  }
  \centerline{
    \plottwo{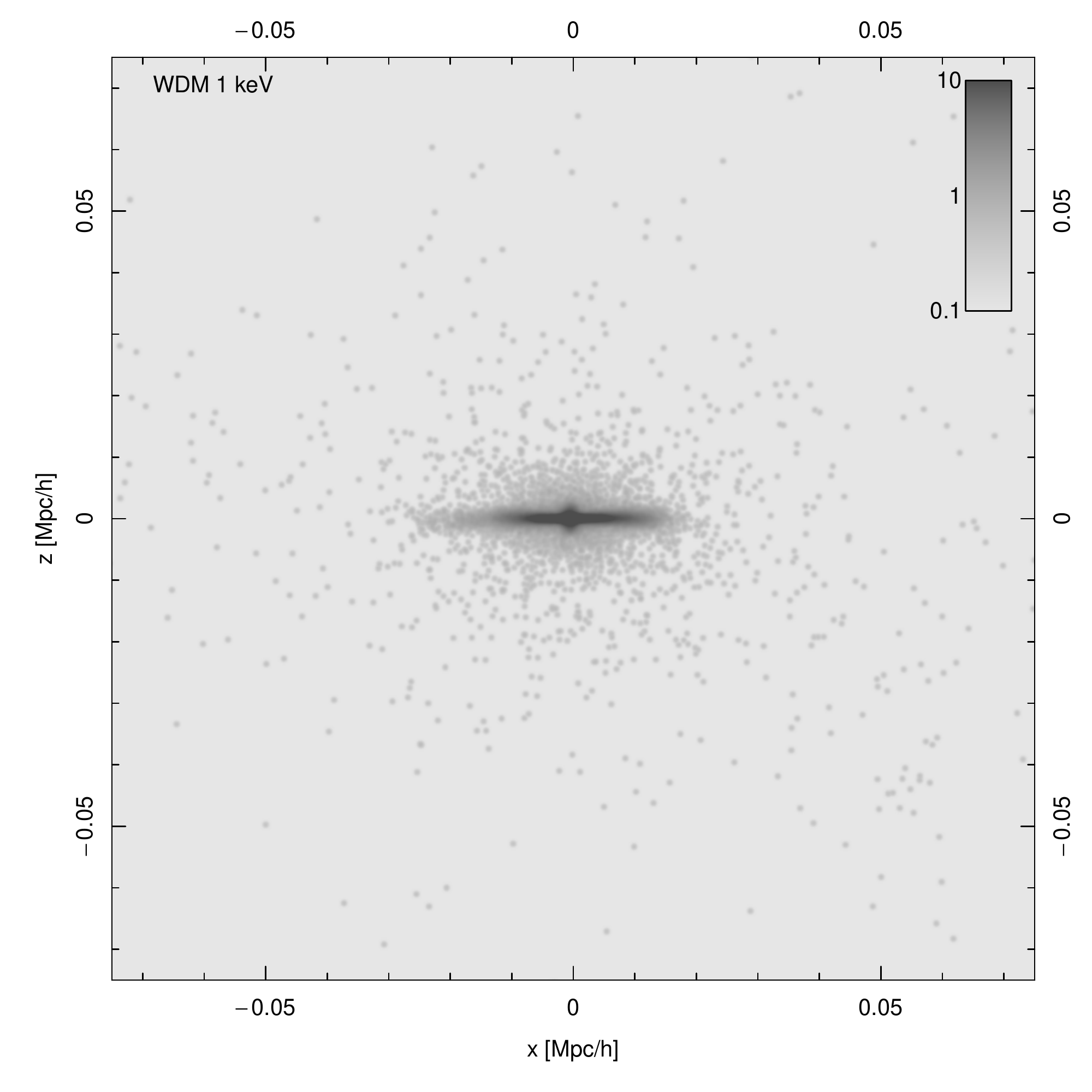}{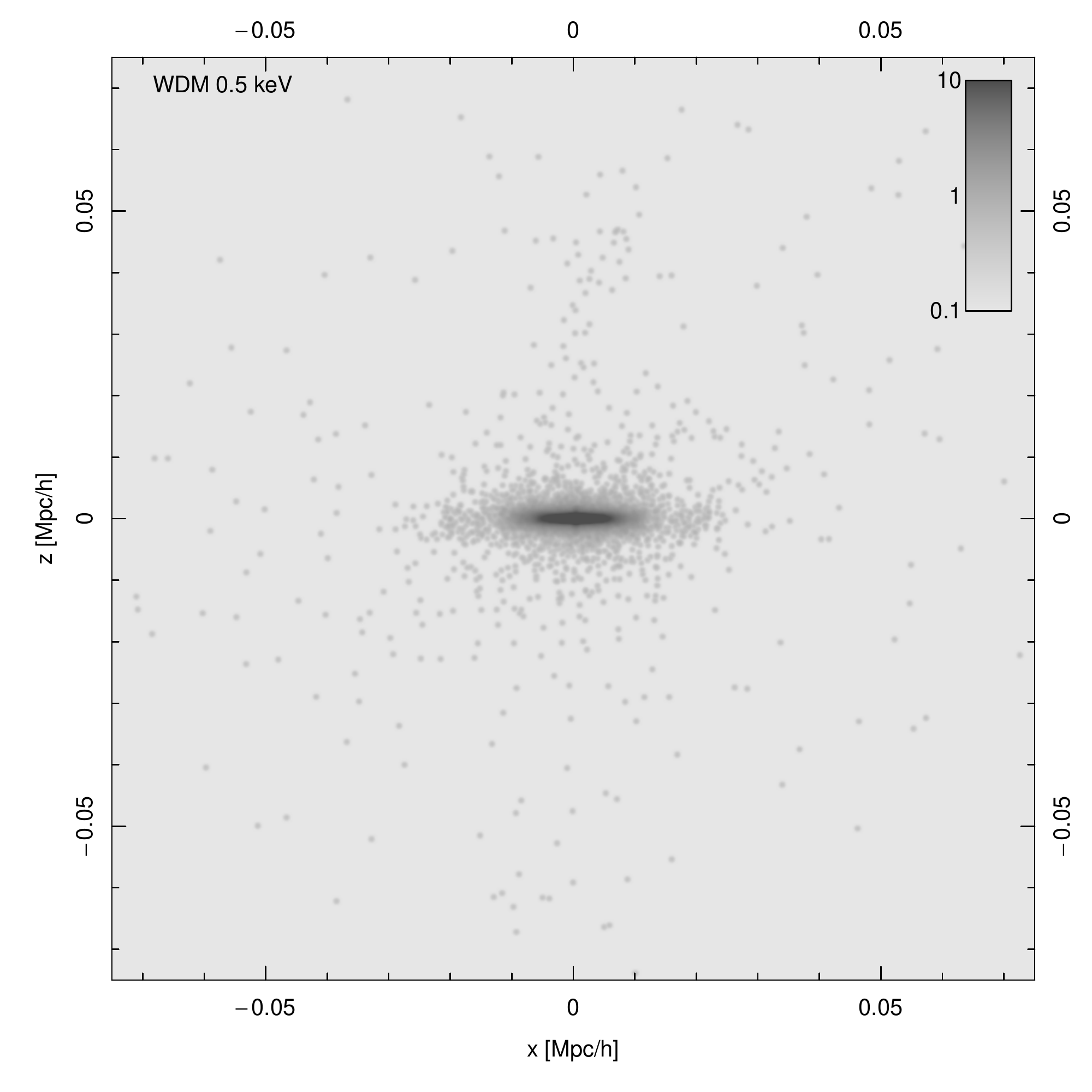}
  }  
  \caption{{\bf Projected density of stellar material within 50 $h^{-1}$ kpc
      radius of galaxy at $z$=0.} The CDM case is shown in the top left;
    the WDM $m_{\rm WDM}$=2, 1, and 0.5 keV/$c^2$ cases are shown in the top right, bottom
    left and right respectively. In each case, we have rotated the galaxy
    coordinates such that it is viewed edge on in the frame of the principal
    axes; see text for further details on how this has been calculated.}
  \label{fig:projected_density_z0}
\end{figure*}


\paragraph*{Visual Impression} In Figure~\ref{fig:projected_density_z0}, we show
how the projected density of
stellar material within a 50 $h^{-1}$ kpc radius, centred on the galaxy, varies
between the fiducial CDM run and its WDM counterparts. In each case, we have
rotated the system such that the central galaxy is viewed edge-on in the
frame of its principal axes, which we do by diagonalising the reduced
moment of inertia tensor computed for star particles within spherical shells,
\begin{equation}
  \label{eq:inertia_tensor}
  I_{ij} = \sum \frac{\Delta x_i \Delta x_j}{r^2}.
\end{equation}
Here we compute offsets in the three spatial components $\Delta x^i$ with respect
to the centre of density of the mass distribution, projected onto the unit
sphere, and then diagonalise to obtain eigenvalues, which provide a measure
of the shape of the stellar mass distribution within the shell, and
eigenvectors, which provide us with the necessary rotation matrix to convert
coordinates into the principal axes. When rotating the system, we use the
rotation matrix derived from the innermost shell of particles.

This Figure reveals a number of points worthy of note. First, the stellar
distribution within the galaxies takes the form of a disc, with a radial
scale length that is broadly the same in each of the runs; visually, the
CDM disc appears slightly more compact than the corresponding WDM discs.
Second, the projected density of extra-disc stellar material is greatest at
small galacto-centric radii, and it is flattened in the direction perpendicular
to the disc. Third, and most significantly for this study, we observe that the 
spatial extent of extra-disc material is similar in the CDM and WDM
$m_{\rm WDM}$=2 keV/$c^2$ runs,
but the density appears to decline more rapidly in the WDM $m_{\rm WDM}$=0.5 and 1
keV$/c^2$ runs.

\medskip

\begin{figure}
  \centerline{\plotone{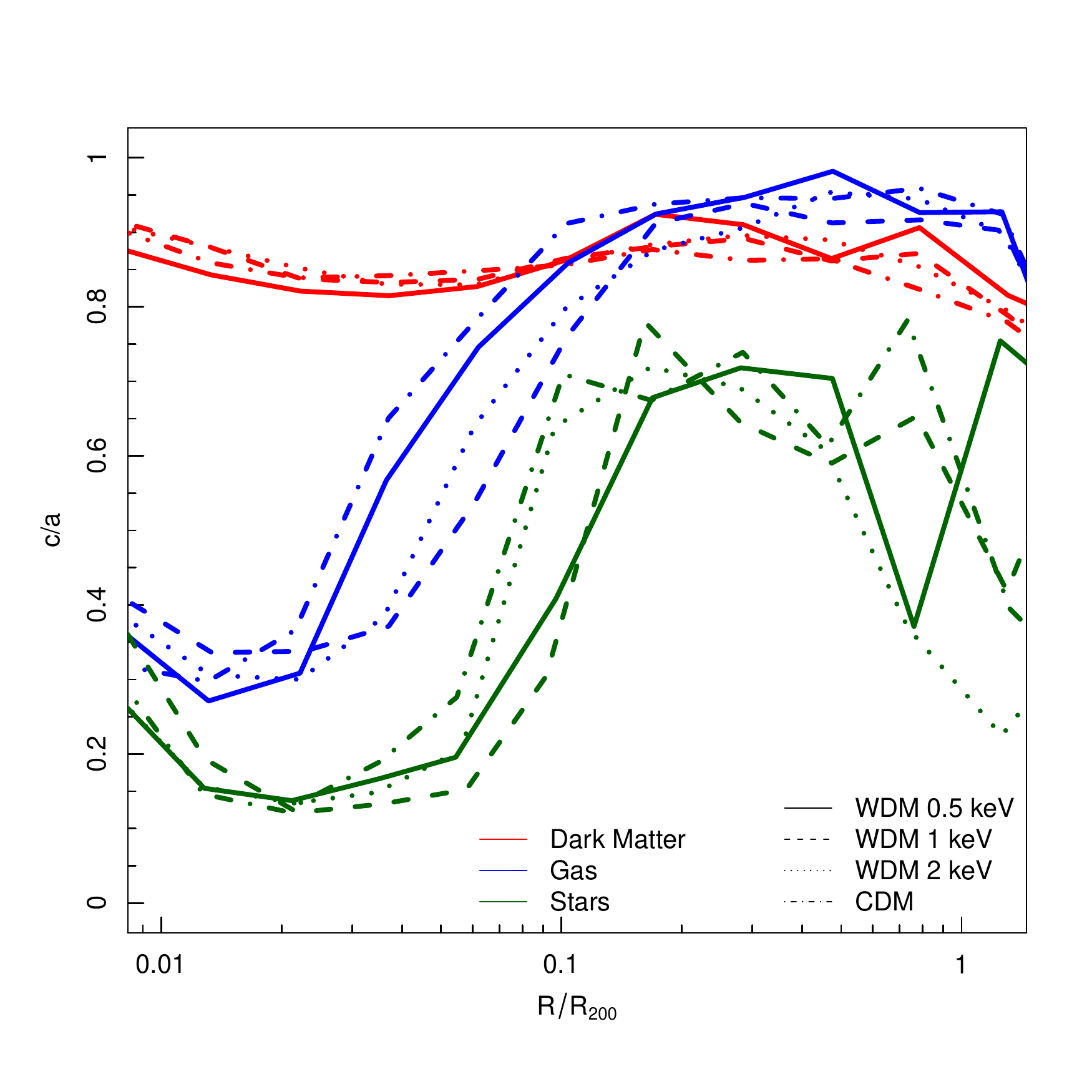}}  
  \caption{{\bf Radial variation of minor-to-major axis (c/a) ratio at $z$=0.}
    We have computed a simple measure of the flattening of the mass
    distribution, as quantified by $c/a$, as a function of radius (see text
    for details). Different curves correspond to stellar, gas, and dark matter
    components (green, blue and red curves) in the fiducial CDM run
    (dotted-dashed curves) and WDM $m_{\rm WDM}$=2, 1, and 0.5 keV/$c^2$ curves
    (dotted, dashed, and solid curves).}
  \label{fig:minor_axes_z0}
\end{figure}

\paragraph*{Mass Distribution}
The flattening evident in Figure~\ref{fig:projected_density_z0} is quantified
in Figure~\ref{fig:minor_axes_z0}, where we show how the minor-to-major axis
ratio $c/a$ varies with radius. The behaviour of $c/a$ measured for the different
components is broadly similar across the different dark matter models --
the dark matter $c/a$ varies little with radius ($\sim 0.8-0.9$),
whereas $c/a$ for both the gas and stars is small at small radii
($\sim 0.2-0.4$ at $R \lesssim 0.05-0.1 R_{200}$) before rising
rapidly between $\sim 0.1-0.2 R_{200}$ to $c/a \sim 0.7$ for the stellar
component and $c/a \sim 0.8-0.9$ for the gas component. This is consistent
with our visual impression in Figure~\ref{fig:projected_density_z0}, at least
for the stellar component.

\medskip

\begin{figure}
  \centerline{\plotone{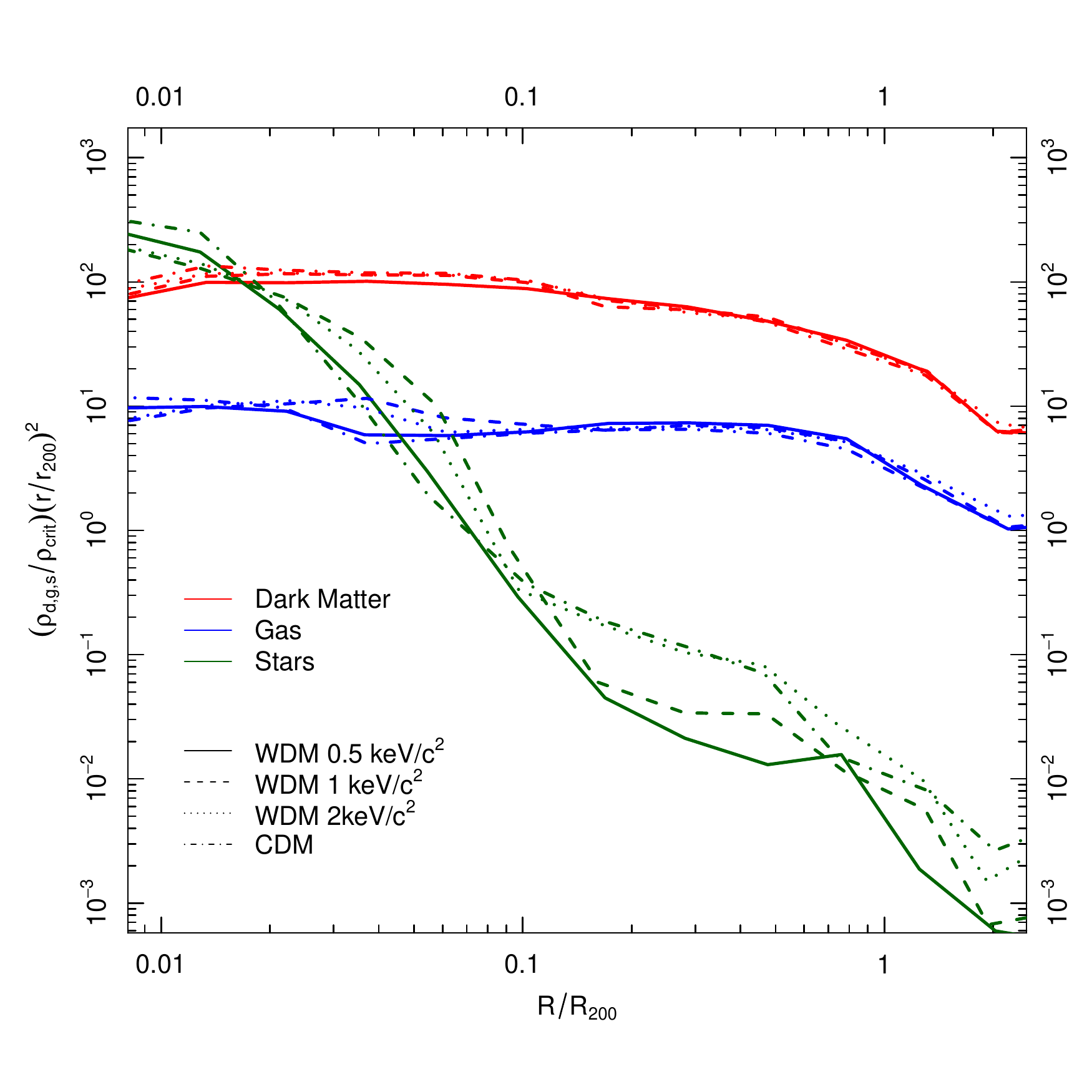}}  
  \caption{{\bf Spherically Averaged Mass Profiles at $z$=0.} Here we show
    spherically averaged stellar, gas, and dark matter mass profiles (green,
    blue and red curves) in the fiducial CDM run (dotted-dashed curves) and
  WDM 2, 1 and 0.5 keV curves (dotted, dashed, and solid curves).}
  \label{fig:density_profile_z0}
\end{figure}

In Figure~\ref{fig:density_profile_z0}, we focus on the spherically averaged
mass density profiles at $z$=0 for the stars, gas, and dark matter
(green, blue, and red curves) in the CDM and WDM $m_{\rm WDM}$=2, 1, and
0.5 keV/$c^2$ keV runs
(dotted-dashed, dotted, dashed, and solid curves) respectively. Profiles
are constructed by defining the centre of density $\vec{r}_{\rm cen}$ of the
composite system using the shrinking spheres method \citep[cf.][]{power.etal.2003},
sorting particles by radius, and assigning them to 25 spherical
logarithmic bins equally spaced between $R_{\rm min}=0.01 R_{200}$ and
$R_{200}$.

Interestingly, differences in the dark matter and gas mass profiles are small
between dark matter models, especially at radii $R/R_{200}\gtrsim 0.1$. At smaller
radii, the spherically averaged gas density differs by a factor of a
few in density over a factor of a few in radius within $R/R_{200}\sim 0.1$,
but these differences occur in the peripheries of the galaxy disc. However, there
are striking differences in the spherically averaged stellar mass density;
outside of the region within which the galaxy disc reside, between
$R/R_{200}\sim 0.1$ and $R/R_{200}\sim 0.5$, we find that the spherically
averaged stellar density is approximately an order of magnitude smaller in
the WDM $m_{\rm WDM}$=1 and 0.5 keV$/c^2$ runs than in the CDM and WDM 2 keV$/c^2$ runs.

As we demonstrate in Appendix Figure~\ref{fig:density_profile_cdm_z0}, we
expect these differences in the
spherically averaged mass profiles of the ESC to be relatively insensitive to our
choice of galaxy parameters parameters -- the properties of the central galaxy
show a much greater dependence on what we assume for the threshold for star formation
($n_{\rm thresh}$) or strength of supernova feedback ($\epsilon_{\rm feed}$). On the
other hand, we expect our results to be sensitive to mass resolution -- as we resolve
lower mass systems and gas can reach higher densities, where star formation occurs and
how it is affected by, for example, feedback, will affect when stars form and the rate
at which satellites disrupt; this is evident as we go to higher resolution. As we show
in Figure \ref{fig:density_profile_cdm_z0}), this is apparent especially within the central
regions where the galaxy resides at $z\simeq 3$ (the latest time at which we we have data
available currently), although the properties of the ESC are reasonably consistent between
resolutions. However, we focus on relative
differences in this study, and we are confident that the runs we use are adequate for
this purpose.

\medskip

\begin{figure}
  \centering
  \plotone{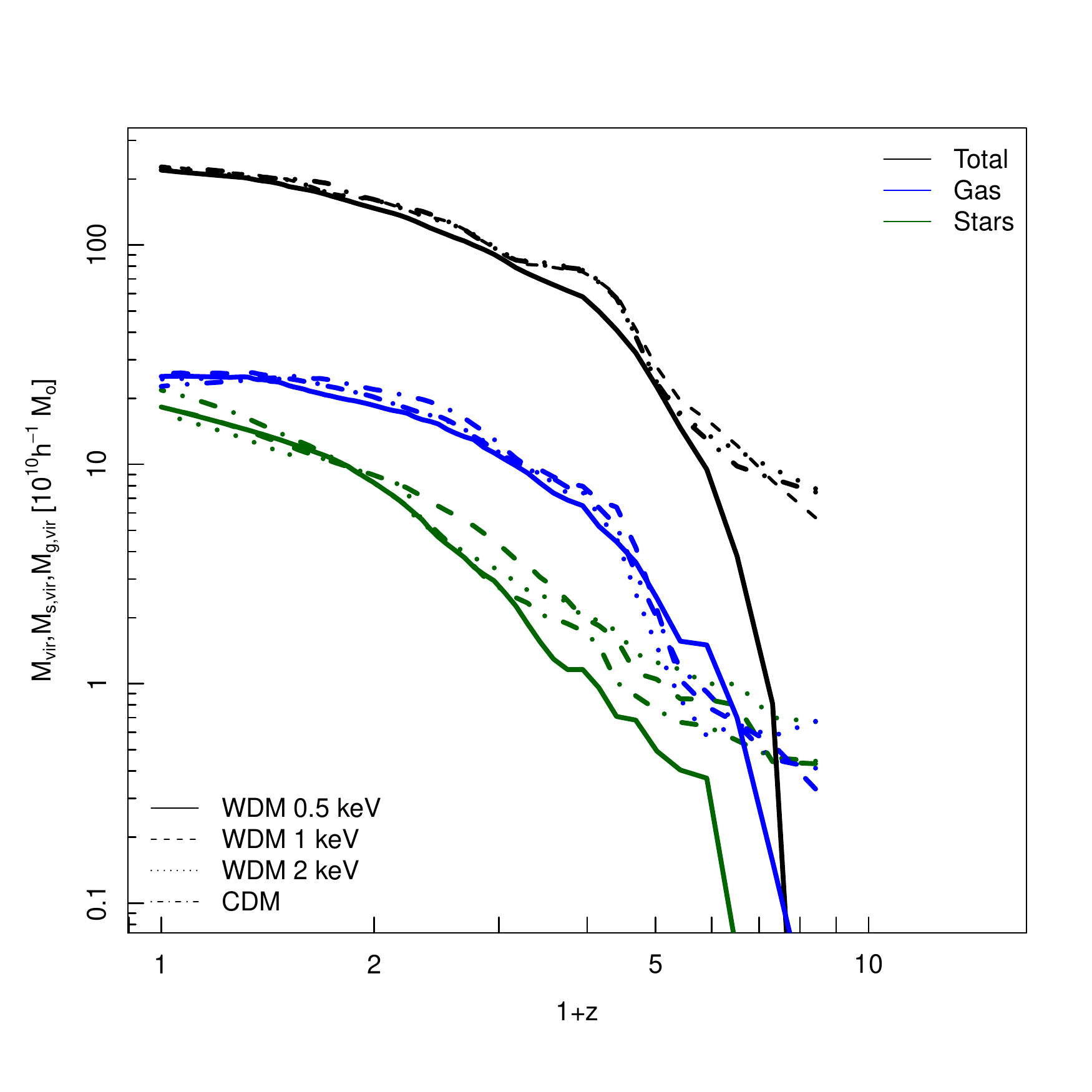}
  \caption{{\bf Growth of Virial Mass with Redshift.} Here we show how
    the virial mass $M_{200}$, including stellar, gas, and dark matter mass
    (black curves), and the stellar and gas masses (green and blue curves
    resectively) within $R_{200}$ have been assembled as a function of
    redshift, in the fiducial CDM model (dotted-dashed curves) and in the
    WDM 2, 1 and 0.5 keV models (dotted, dashed, and solid curves
    respectively).}
  \label{fig:assembly_history_mvir}
\end{figure}

To what extent are these differences at $z$=0 evident at earlier times?
In Figure~\ref{fig:assembly_history_mvir}, we track the growth of the total
virial mass, which we define as $M_{200}$ at all redshifts, as well as the gas
and stellar masses within $R_{200}$ at each redshift. At late times, the
masses in the different components are similar across the models, but as we
go to earlier times ($z\gtrsim 4$), differences between the models become
apparent. The rate at which both the total mass (upper curves) and gas
mass (middle curves) grows is similar in the CDM, WDM 2 and 1 keV models,
but there is a substantial lag in the 0.5 keV model; the rate at which the
stellar mass grows shows greater variation between the models and is evident
down as late as $z \sim 1$.

\begin{figure}
  \centering
  \plotone{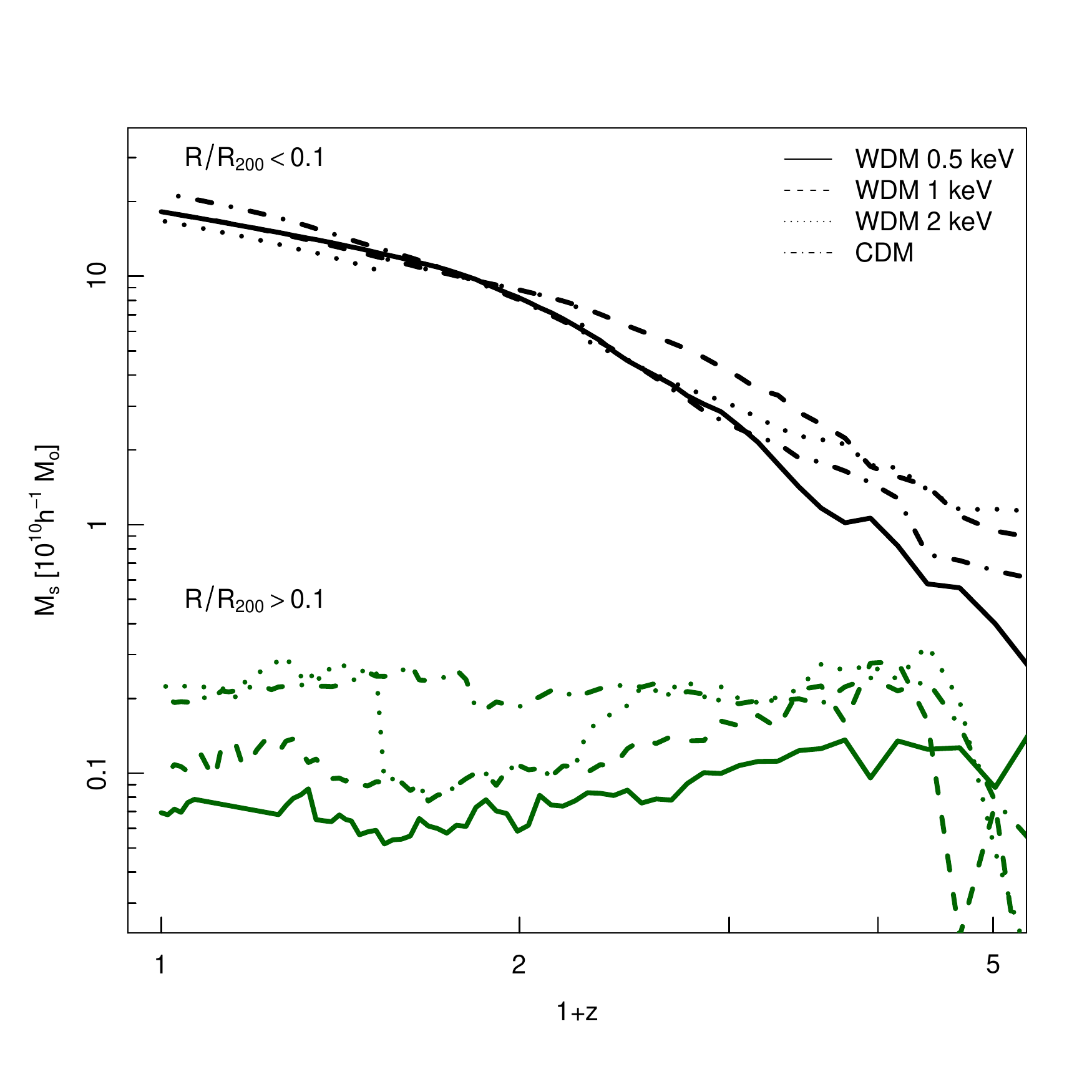}
  \caption{{\bf Growth of Stellar Mass in Halo with Redshift.} Here we
    show how the stellar mass in the halo between $0.1 \leq R/R_{200} \leq 0.8$,
    in the fiducial CDM model (dotted-dashed curves) and in the
    WDM 2, 1 and 0.5 keV models (dotted, dashed, and solid curves).}
  \label{fig:assembly_stellar_mass}
\end{figure}

We explore the growth the stellar mass in more detail in
Figure~\ref{fig:assembly_stellar_mass}, where we separate stellar mass within
$R_{200}$ into contributions from $R/R_{200}<0.1$, which contain the central
galaxy (upper curves), and $0.1 \leq R/R_{200} \leq 0.8$ (lower curves). This
reveals that, at late times especially, the amount of stellar mass associated
with the central galaxy is consistent between the different dark matter runs;
the differences that we see in Figure~\ref{fig:assembly_history_mvir} are
driven by differences in the outer stellar component surrounding the galaxy.
The contribution from the outer stellar component in the CDM model is
$\sim 2 \times 10^9 h^{-1} \rm M_{\odot}$ at $z$=0 and has remained at this level
since $z \simeq 4$; the contributions in the WDM 1 and 0.5 keV models are 
approximately 1/2 and 1/4 of this at $z$=0, but these were larger in the
past, by roughly a factor of 2 at $z\simeq 4$.

\begin{figure}
  \centering
  \plotone{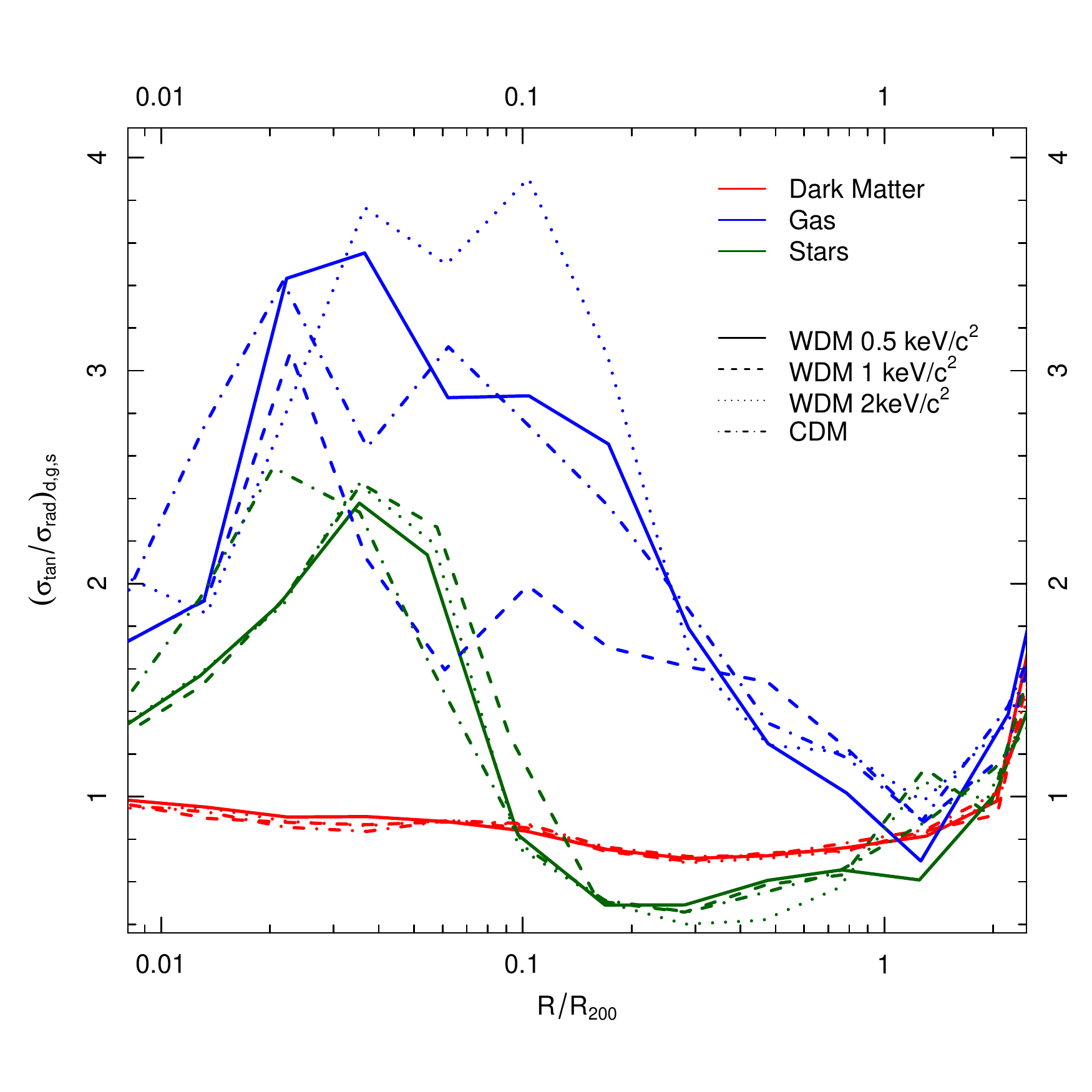}
  \caption{{\bf An Estimate of the Velocity Anisotropy at $z$=0.} Here we
    estimate the relative importance of radial to tangential motions by
    showing the ratio of the spherically averaged stellar, gas, and dark
    matter tangential and radial velocity dispersions (green, blue, and
    red curves) in the fiducial CDM (dotted-dashed curves) and WDM 2, 1, and
    0.5 keV (dotted, dashed, and solid curves) runs.}
  \label{fig:velocity_anisotropy_z0}
\end{figure}

\medskip
\paragraph*{Kinematics and Orbits}
Figures~\ref{fig:density_profile_z0} and~\ref{fig:assembly_stellar_mass}
confirm our visual impression that the density of stellar material outside of the
galaxy disc drops off more rapidly in the WDM 1 and 0.5 keV runs when compared to
the CDM and WDM 2 keV runs. In 
Figure~\ref{fig:velocity_anisotropy_z0} we investigate whether or not this
reflects differences in the kinematics of the stellar material, because the
nature of the orbits that a population of stars follow will be imprinted on the mass
density profile. We quantify this by considering the ratio of the tangential
and radial velocity dispersions, $\sigma_{\rm tan}$ and $\sigma_{\rm rad}$,
which allow us to estimate the velocity anisotropy as a function of radius.
$\sigma_{\rm tan}/\sigma_{\rm rad}\sim 1$ indicates that orbits are approximately
isotropic; $\sigma_{\rm tan}/\sigma_{\rm rad} < 1$ indicates that orbits are
preferentially radial; while $\sigma_{\rm tan}/\sigma_{\rm rad} > 1$ indicates
preferentially tangential orbits.

Figure~\ref{fig:velocity_anisotropy_z0} reveals that the velocity anistropy
of the dark matter is similar in each of the runs, isotropic at small radii
and becoming mildly radial at larger radii. The velocity anisotropy of the
gas shows that it is preferentially tangential (unsurprisingly) within
$R_{200}$. Between $R/R_{200} \sim 0.3$ and $R/R_{200} \sim 1$, it is mildly
tangential and declining with increasing radius, and the behaviour is broadly
similar between the runs. Within $R/R_{200} \sim 0.3$, the behaviour differs
sharply and with no obvious trend with underlying dark matter model; the
anisotropy peaks with $\sigma_{\rm tan}/\sigma_{\rm rad} \sim 3-4$ between
$R/R_{200} \sim 0.02-0.1$. The stellar material follows preferentially
tangetial orbits within $R/R_{200}\sim 0.1$, as we would expect from the
degree of flattening evident in Figure~\ref{fig:projected_density_z0}, but
becomes markedly radial between $R/R_{200}\sim 0.1$ and $R/R_{200}\sim 1$.
There are differences within $R/R_{200}\sim 0.1$ between the CDM run and the
WDM runs; the anistropy profile peaks and starts to roll over at a smaller
radius ($R/R_{200}\sim 0.02$) in the CDM run than in the WDM runs
($R/R_{200}\sim 0.04$), which reflects the slightly smaller radial scale
length we have observed already. At the larger radii of most interest when
considering the ESC, however, we find no appreciable differences.

\medskip

\begin{figure*}
  \centerline{
    \plottwo{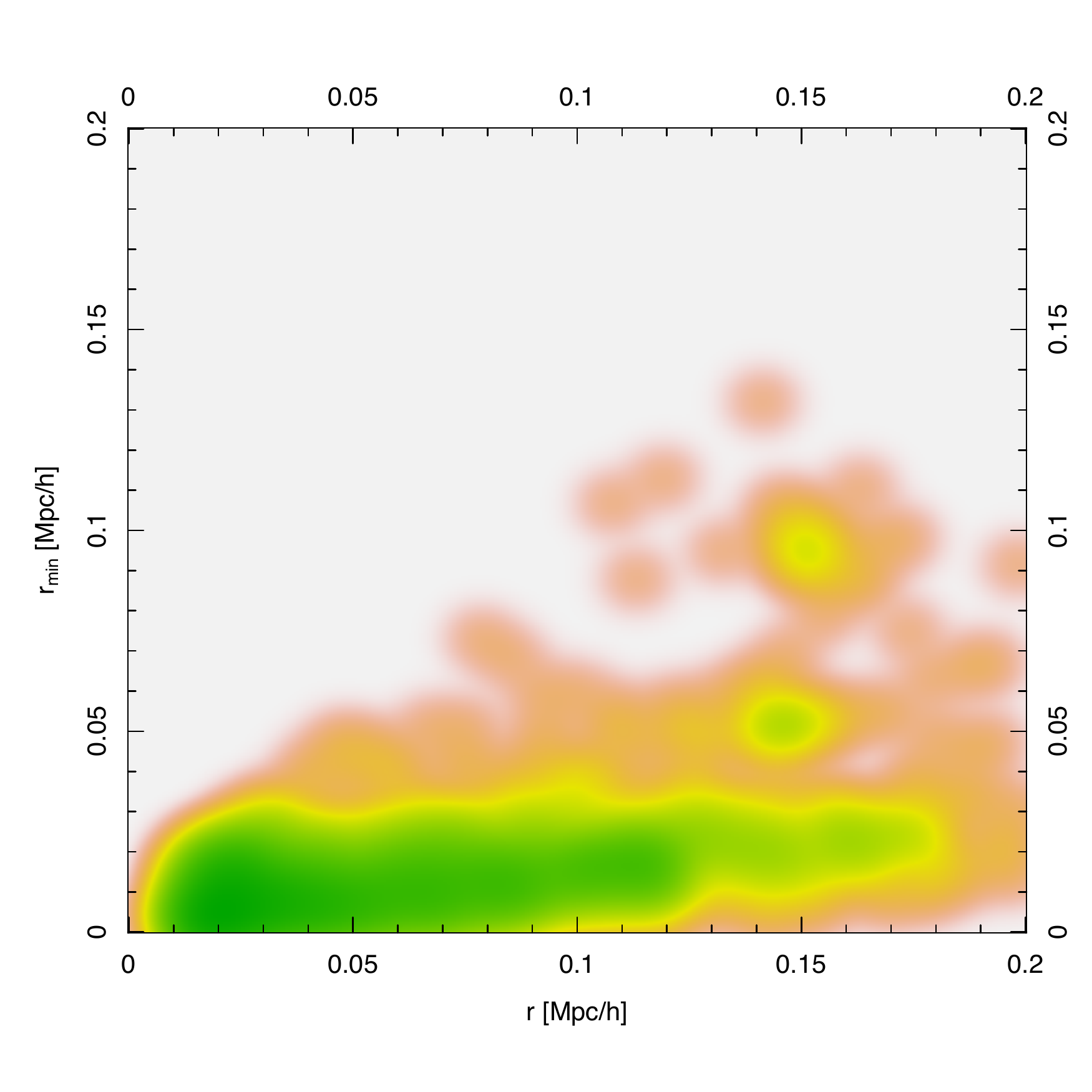}{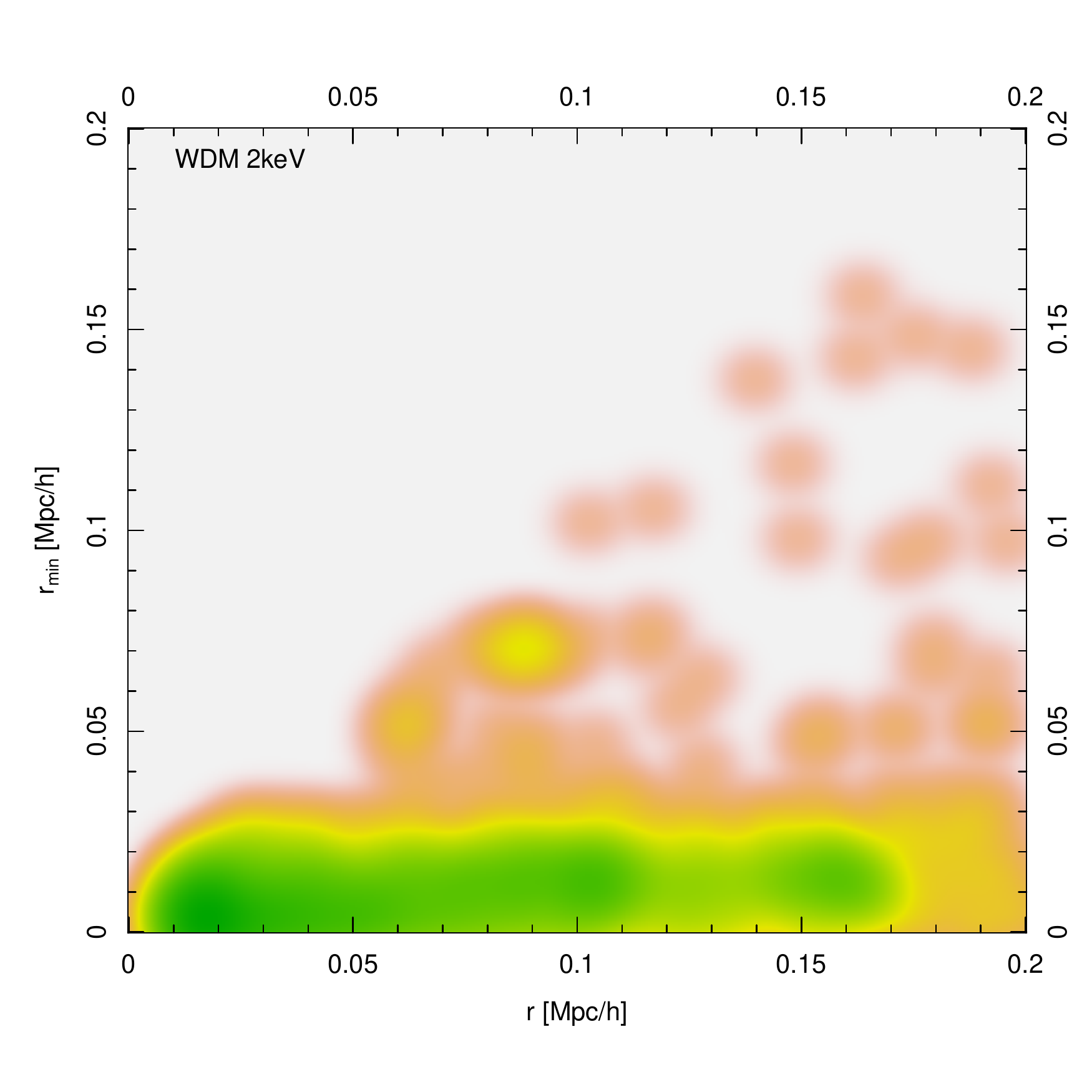}
  }
  \centerline{
    \plottwo{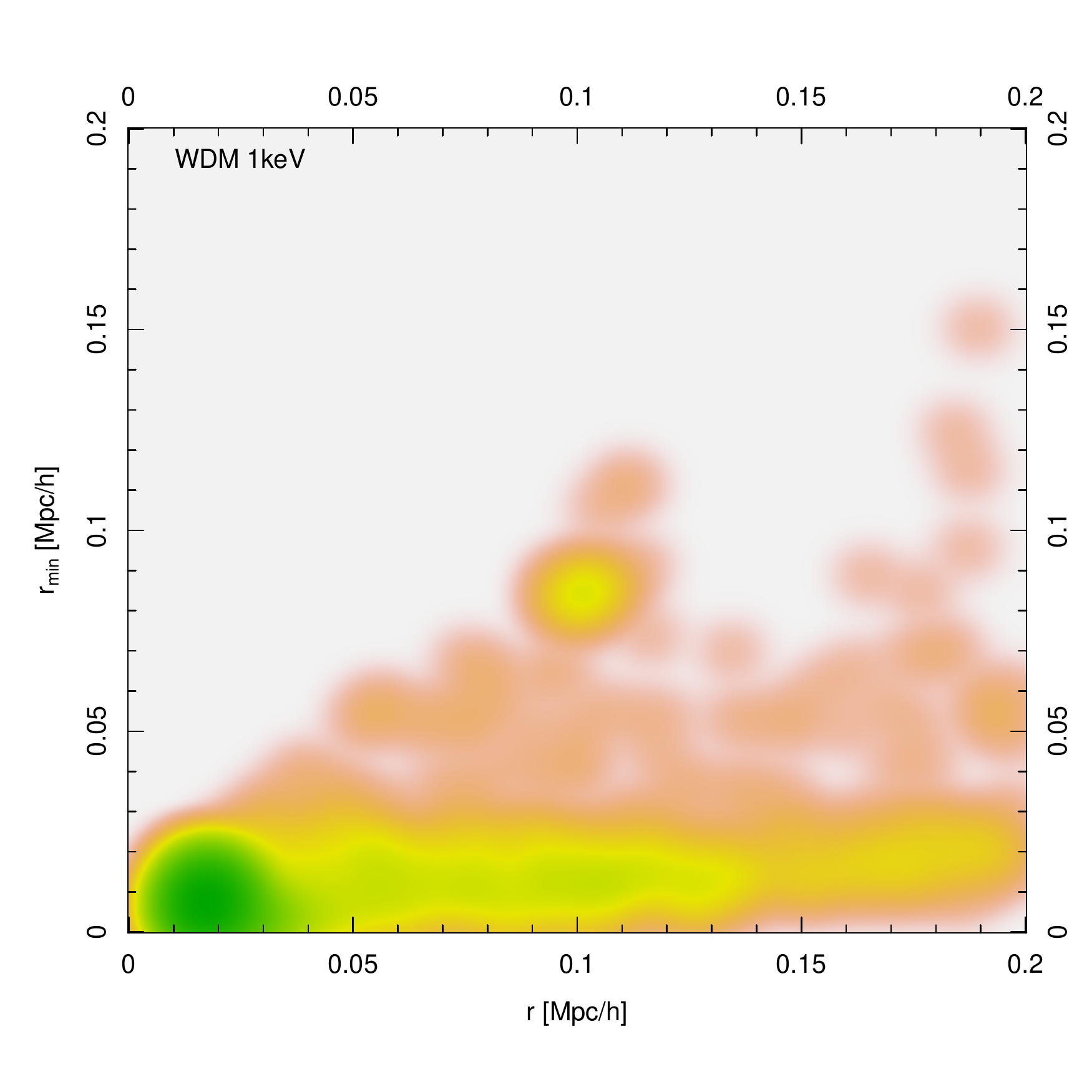}{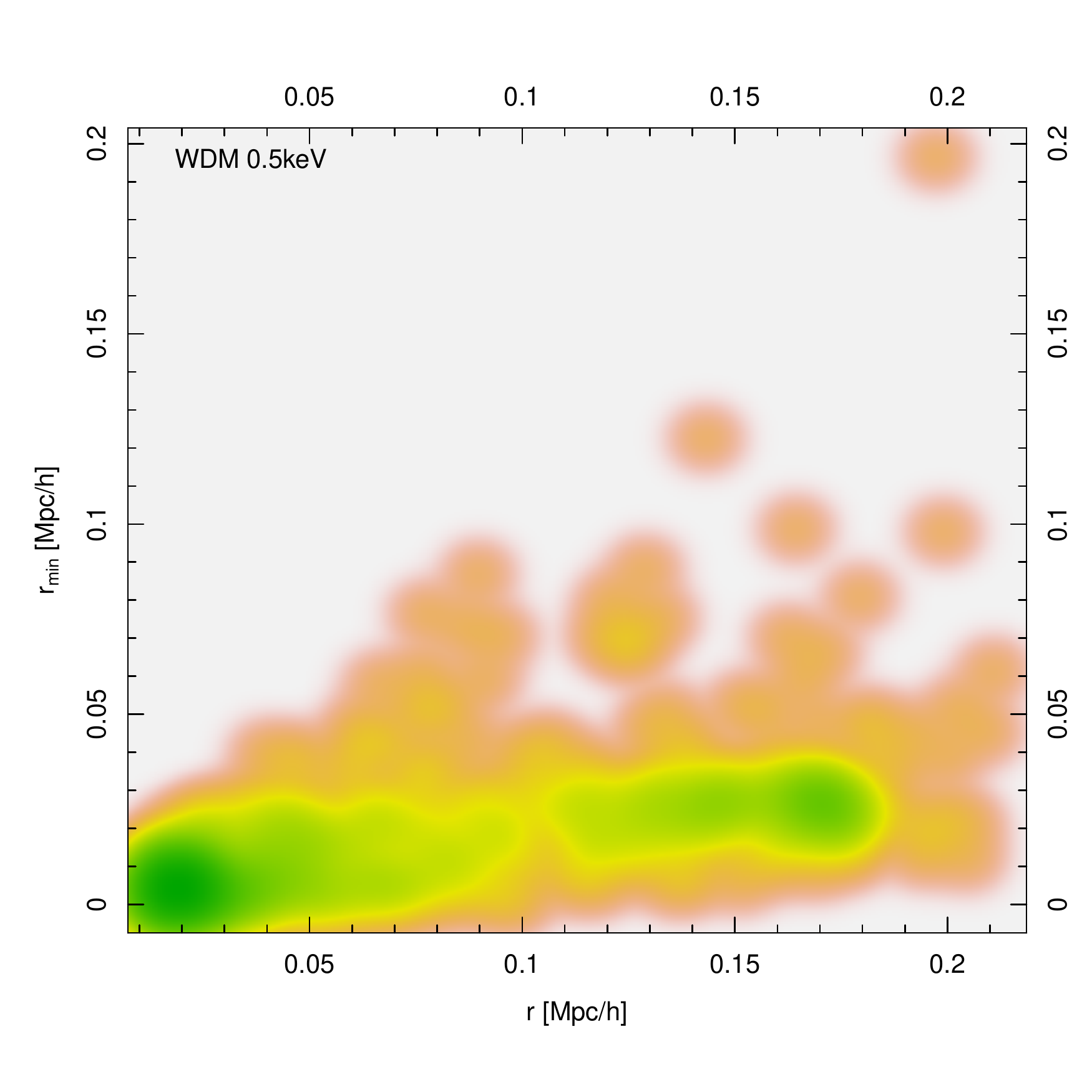}
  }  
  \caption{{\bf Orbital eccentricities of the extended stellar component} Here
    we have tracked
    the orbits of individual star particles and estimated the eccentricity $e$ as the
    ratio of $r_{\rm min}$ to $r_{\rm max}$, where where both are computed following
    initial pericentric passage. The CDM case is shown in the top left; the WDM 2 keV,
    1 keV and 0.5 keV cases are shown in the top right, bottom left and right
    respectively.}
  \label{fig:orbital_eccentricities}
\end{figure*}

Could differences in the spatial structure of the ESC reflect differences
in the kinds of orbits traced out by the progenitors of the material that composes
in the ESC? We check this explicitly in Figure~\ref{fig:orbital_eccentricities},
where we probe the orbits of
the stellar material that lie within the radial range $0.1 \lesssim R/R_{200} \leq 1$
at $z$=0. Here we characterise the orbit by $R_{\rm min}$, the minimum peri-centric
distance from the centre of the galaxy, estimated from the minimum turning point of
the curve tracking star particle radius versus time\footnote{Note that this is
  distinct from pericentre of first infall; this minimum decreases gradually with
  the number of orbits.}; star particles that are on their initial infall onto the
system lie on the diagonal of $R_{\rm min}$ versus $R_0$, where
$R_0$ is the present day radius. In this Figure, we smooth the distribution
of $R_{\rm min}$ and $R_0$ using a 2D binned kernel density estimate. Our expectation
is that satellite galaxies will follow preferentially radial orbits that bring
them close to the centre of the potential, and this is borne out by this plot --
the majority of the stellar material has values of $R_{\rm min}$ within a narrow
range between $\sim 10$ and $\sim 20 h^{-1} {\rm kpc}$. Few of the star particles
that constitute the ESC are infalling for the first time, and satellites that have
yet to disrupt are apparent as dense knots in the distribution. The key point here
is that there is no systematic difference between the distributions in the
four models; orbits are preferentially radial (cf. \citealt{benson.2005},
\citealt{abadi.etal.2006}, \citealt{khochfar.burkert.2006}, \citealt{font.etal.2011},
\citealt{wetzel.2011}, \citealt{rashkov.etal.2013}, \citealt{jiang.etal.2015}) and there is no
compelling reason why the distribution should differ between
CDM and plausible WDM models, as is borne out by numerical simulations
\citep[e.g.][]{knebe.etal.2008} -- material funnels from the cosmic web and into
the potential of the halo, which does not differ substantially between models.

\paragraph*{A Potential Test of Dark Matter?} The results presented so far suggest
that the low surface brightness surroundings of galaxies could contain the observable
imprint of dark matter. Could, because the differences between the fiducial CDM model
and WDM are really
only evident in the runs with the more extreme candidates
(i.e. $m_{\rm WDM}<$2 keV/$c^2$). However, for the same reasons that we argued
that there should be differences betweenn the CDM model and dark matter models
in which the abundance of substructure is suppressed, we can also argue that
there should be natural halo-to-halo variation within the CDM model, reflecting
variations in assembly histories.

\begin{figure}
  \plotone{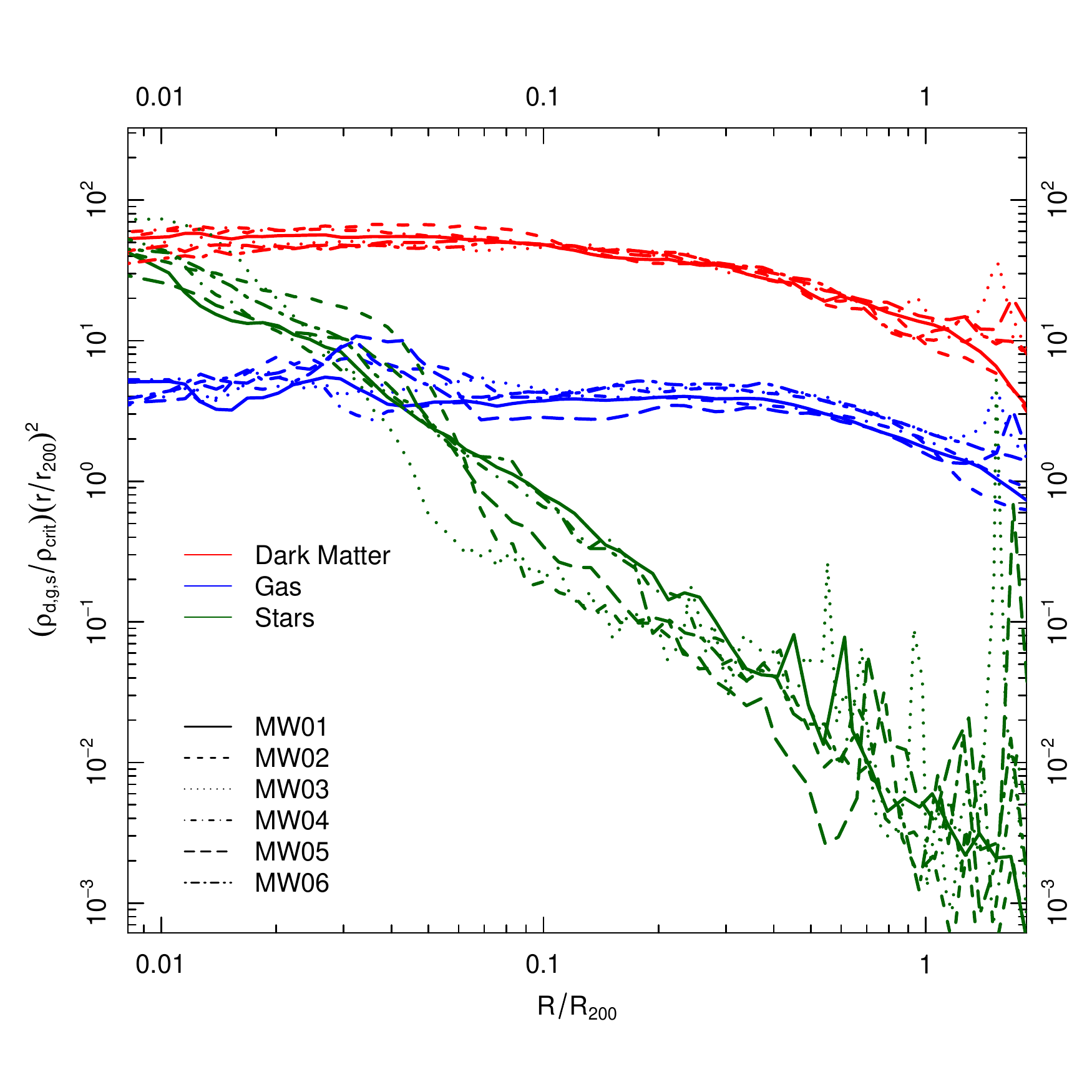}
  \caption{{\bf Galaxy-to-galaxy variation in the CDM model.} Here we show
    how much variation between different galaxies we might expect by plotting
    spherically averaged stellar, gas, and dark matter mass profiles (green,
    blue and red curves) for the CDM run already analysed and a further five
    galaxies, all selected to form in low-density regions with
    $M_{200}\simeq 2 \times 10^{12} h^{-1} {\rm M}_{\odot}$ at $z$=0.}
  \label{fig:density_profile_variation_z0}
\end{figure}

In Figure~\ref{fig:density_profile_variation_z0}, we show spherically averaged density
profiles for six runs in the CDM model (including the run already presented), all with
virial masses of $M_{200} \simeq 2 \times 10^{12} h^{-1} {\rm M}_{\odot}$ at $z$=0
(cf. Table~\ref{tab:assembly_histories}), and all selected to lie in low-density
environments. This Figure highlights the difficulty of using the ESC as a test
of dark matter -- the system-to-system variation in the properties of the ESC are as
large as we see in the model-to-model variation when we vary the underlying dark matter
model. However, it does suggest that the ESC can be used to extend the
concept of galactic archaeology to systems beyond the Milky Way and Andromeda, as is
being done in nearby galaxies \citep[e.g.][]{martinez-delgado.etal.2010,radburn-smith.etal.2011,dragonfly}
and groups \citep[e.g. the PISCeS survey][]{crnojevic.etal.2015}, and will become possible for
statistical samples of galaxies within the Local Volume with, for example, {\small LSST} \citep[cf.][]{lsst}.
Combining metallicity, kinematics, and spatial structure, it should be possible to
trace the assembly history of galaxies.

\section{Summary}
\label{sec:summary}
We have used a set of cosmological zoom galaxy formation simulations of
Milky Way mass galaxies, with $M_{200} \simeq 2 \times 10^{12} h^{-1}
{\rm M}_{\odot}$ at $z$=0, to explore whether or not the spatial and kinematic
properties of the diffuse, extended stellar components (ESCs), in which galaxies
are embedded, might depend on the underlying dark matter model. In our simulations
of a single system in which we vary the underlying dark matter model, this ESC extends
from approximately $15 h^{-1} {\rm kpc}$, the outskirts of the
galaxy disc, to $R_{200} \simeq 200 h^{-1} {\rm kpc}$, and the stellar remnants of
disrupting satellite galaxies make a signficant contribution to its mass. For our
dark matter models, we considered fiducial Cold Dark Matter (CDM) and Warm Dark
Matter (WDM) alternatives with particle masses of $m_{\rm WDM}$=0.5, 1, and 2
keV/$c^2$; as we discussed in \S\ref{sec:sims}, although models with
$m_{\rm WDM}<2 {\rm keV}/c^2$ are not favoured by current observational limits,
we are interested in establishing whether or not dark matter models that alter
the abundance of substructure could leave an imprint on observable properties of
galaxies. Because the orbital properties of subhalos in CDM and WDM models are
similar, the suppression of low-mass subhalos in WDM models means that the average
subhalo mass is more massive than in the corresponding CDM model, which implies
shorter merging timescales on average. The tail of low-mass subhalos with long
merging timescales in the CDM model means that satellites disrupting in the tidal
field of the galaxy lose mass over an extended radial range, tracking their orbit.
In other words, the nature of
dark matter should be evident in the structure of the ESC, which should be more
centrally concentrated in WDM models with lower particle masses (i.e. warmer dark matter).

Our results confirm our expectation that properties of the ESC do vary with
the underlying dark matter model, but that differences are most readily
apparent only in the more exteme WDM models that we consider, with $m_{\rm WDM}<2 {\rm keV}/c^2$.
The average stellar density between $15 h^{-1} {\rm kpc} \lesssim R \lesssim 150 h^{-1}
{\rm kpc}$ in the $m_{\rm WDM}$=0.5 and 1 keV/$c^2$ runs is a factor of $\sim 10$
smaller than in the $m_{\rm WDM}$=2 keV/$c^2$ and CDM runs. This difference is
imprinted early in the history of the galaxy at least back to $z \simeq 4$ -- and
persists to the present day. Interestingly, the properties of the central galaxy, a
thin rotationally supported disc, are in good agreement between the different models,
consistent with the conclusions of previous studies \citep[e.g.][]{herpich.etal.2014}.
Otherwise, the degree of flattening as measured by the axis
ratio $c/a$; the kinematics as measured by the ratio of radial to tangential velocity
dispersions, $\sigma_{\rm tan}/\sigma_{\rm rad}$; and the orbits of stars, as estimated
from the distribution of pericentric distance $r_{\rm min}$ to present day position
$r_0$, are all indistinguishable between models. Similarly, the properties of the
dark matter halo and gas are also in very good agreement between models.

Using properties of the ESC to place observational limits on plausible dark matter models
that suppress the abundance of low-mass subhalos, and consequently satellites, is likely
to be challenging, however. Analysing the results of a further five zoom simulations
of Milky Way mass galaxies, all selected to lie within a similar low-density environment,
we find sufficient system-to-system variation in the properties of the spherically averaged
stellar density profile beyond the central stellar component -- all flattened and rotationally
supported -- to make observational measurements using spatial structure alone that distinguish
between even the more extreme WDM models (i.e. $m_{\rm WDM}<2 {\rm keV}/c^2$) and CDM
unconvincing.

Note that we have focussed essentially on the unresolved ESC -- simply considering the
spatial and kinematic distribution of stellar material in galactic outskirts, because these should be
accessible to deep imaging surveys, possibly stacking large numbers of galaxies by central galaxy
stellar mass or halo mass bins. It is worth noting that there may be potential to look at the
resolved ESC and to combine spatial, kinematic and metallicity substructure information to test
dark matter; this will require the kind of statistical sample of Local Volume galaxies that will be
accessible with LSST \citep{lsst}. Even if this remains a challenging test of dark matter, there is
good reason to expect that we can use properties of the ESC to explore the mass assembly histories
of galaxies, tracing merger and accretion events using material in the outer halo, and placing limits
of the growth of galaxies in the context of their larger scale environment. We will explore this
idea in forthcoming work.

\section*{Acknowledgments}
\noindent The authors thank the referee for their careful reading of the paper.
CP acknowledges support of Australian Research Council (ARC) Future
Fellowship FT130100041 and Discovery Project DP130100117. ASGR acknowledges
support of a University of Western Australia Research Fellowship. Both CP and
ASGR acknowledge support of an ARC Discovery Project DP140100198. CP thanks
Alexander Hobbs and Justin Read for making this version of {\small GADGET-SPHS}
available. The research presented in this paper is undertaken as part of
the Survey Simulation Pipeline (SSimPL; {\texttt{http://ssimpl.org/}). This
work was supported by resources provided by the Pawsey Supercomputing
Centre with funding from the Australian Government and the Government of
Western Australia.

\appendix

\section{Sensitivity to Galaxy Formation Parameters}
\label{sec:sens_params}

\noindent Here we explore the extent to which the structure of the extended stellar
component, as well as the central stellar, gas, and dark matter components, as
sensitive to our choice of galaxy formation parameters. We consider variations in three
parameters;
\begin{enumerate}
\item the physical threshold density for star formation, $n_{\rm thresh}$; 
\item the strength of feedback, $\epsilon_{\rm feed}$; and
\item the temperature floor, $T_{\rm floor}$, which is the minimum temperature that we allow gas to cool to. 
\end{enumerate}
In the first two cases, we consider larger and smaller values of the parameters
relative to the fiducial case (i.e. more/less efficient star formation of $n_{\rm thresh}=0.1/100\,n_{\rm thresh}^{\rm fid}$, strong/no feedback, $\epsilon_{\rm feed}=5/0\,\epsilon_{\rm feed}^{\rm fid}$); in the third, we adopt a temperature floor of $T_{\rm floor}$=10$^4$K,
rather than the 100K. In addition, we compare the results of our standard run, using
fiducial galaxy formation parameters, with a particle mass of $1/5^{\rm th}$ finer in mass
resolution. We show the resulting radial density profiles ($\rho\,r^2$) in
Figure~\ref{fig:density_profile_cdm_z0}.

\medskip
In the left hand panel, where we investigate the sensitivity of our results to galaxy
formation parameters,  the most striking differences arise at small
radii, within the galaxy that forms -- increasing the temperature floor, or the strength
of feedback, results in a drop in central density. Interestingly, varying the threshold
for star formation also results in a drop in central density; because $\epsilon_{\rm feed}$ is kept fixed, the combination of more efficient star formation and feedback compared
to inefficient feedback appears to conspire to produce comparable central densities.
As far as the properties of the extended stellar component are affected, all of the
runs bar ``No Feedback'' produce mass profiles that are similar -- the stellar component
has a similar amplitude and shape, while the gas and especially the dark matter
components are very similar. In the absence of feedback, the stellar mass density
over the radial range $\sim 0.05$ to $\sim 1$ $R_{200}$ is factor of $\sim 10$ greater
than in the other runs. The central stellar and gas densities within the galaxy are
similar to the fiducial run, but we note that the central dark matter density is
lowered by a factor of a few.

\medskip
In the right hand panel, we compare results of the system at $z\simeq 3$, the latest time
we have available for the higher resolution run currently. Differences in the central stellar
density is apparent, with higher densities in the lower resolution run, as we would
expect given the relatively coarse resolution of our run \citep[cf.][]{governato.etal.2010}.
However, the stellar density over the radial range we are interested in is broadly consistent.
It remains to be seen whether or not this holds at $z$=0 and at even higher resolution, but
at this point we might expect variations arising from our implementation of the physics
of galaxy formation to be as important, if not moreso.

\begin{figure}
  \plottwo{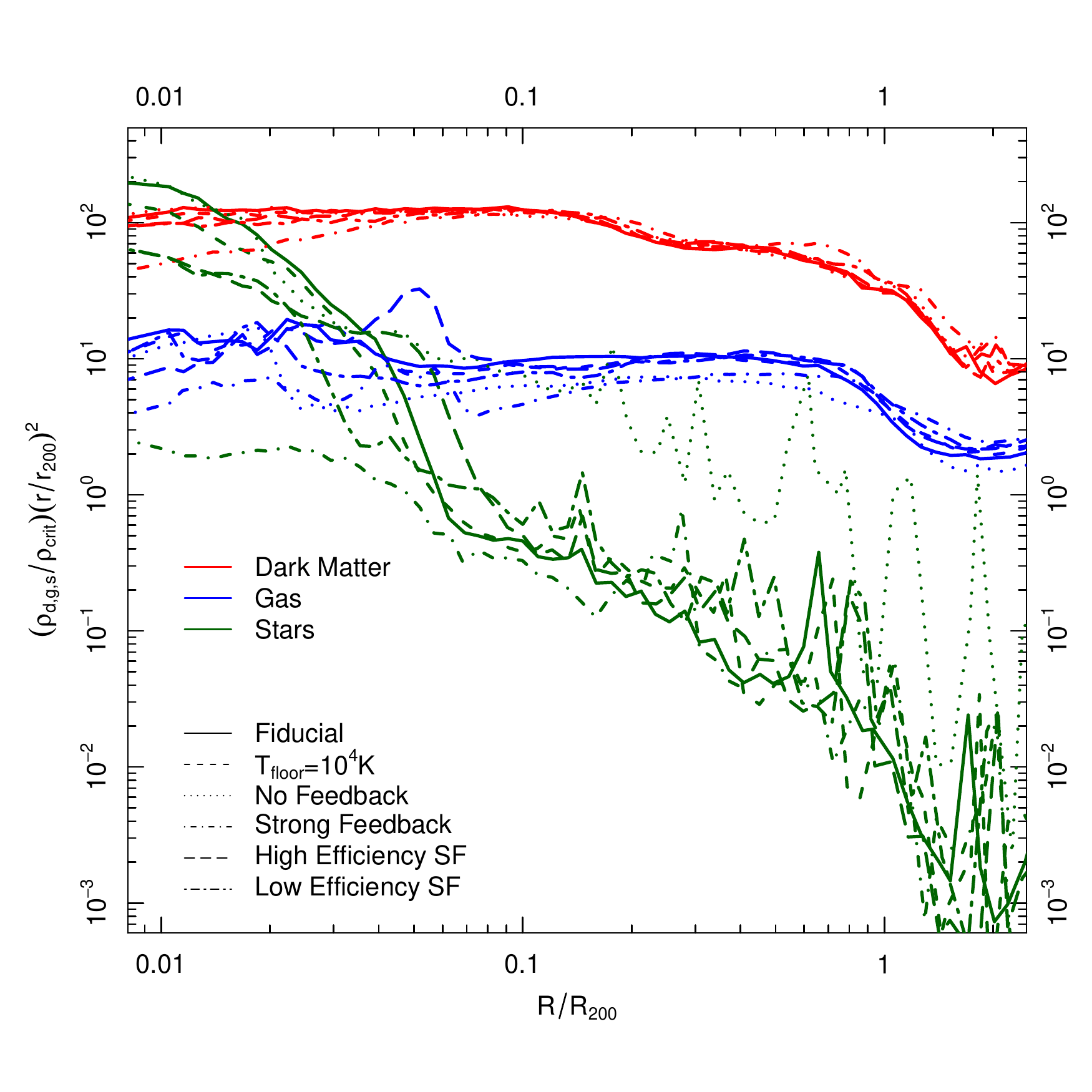}{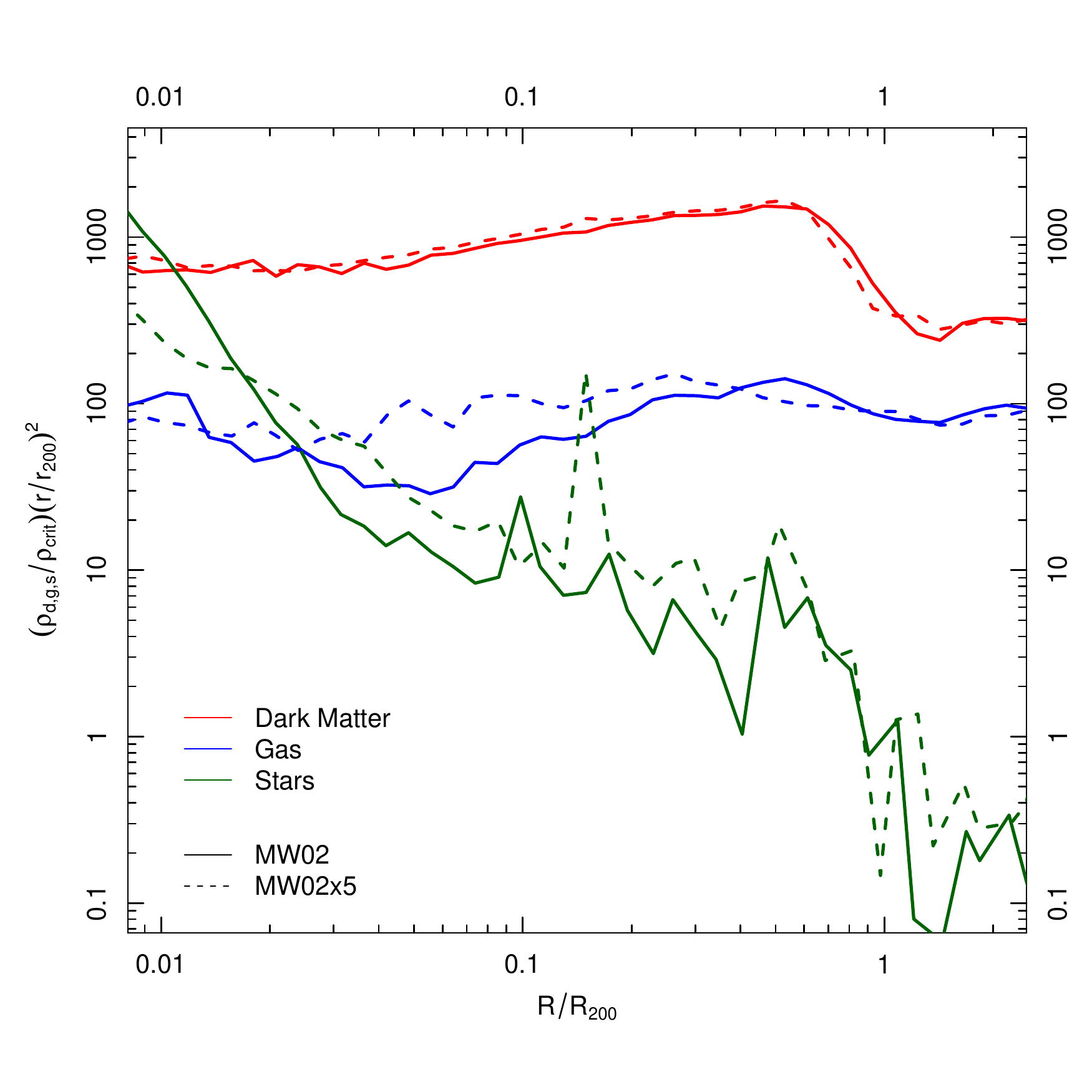}
  \caption{{\bf Influence of Galaxy Formation Parameters and Mass Resolution.}
    {\emph Left:}
    Here we show how our
    choice of galaxy formation parameters affect the spherically averaged stellar,
    gas, and dark matter mass profiles (green, blue and red curves) in the galaxy at $z \simeq 1$. Solid curves
    correspond to our fiducial parameter set, while the dashed curve corresponds to
    a run with an increased temperature floor of 10$^4$K. The remaining curves
    indicate runs with strong and no feedback (dotted and dotted-dashed), and
    more and less efficient star formation (long dashed and short-dashed-long dashed.
    {\emph Right:} Here we show how mass resolution affects spherically averaged stellar,
    at $z\simeq 3$; fiducial galaxy formation parameters are used. The high resolution
    run (dashed curves) has a particle mass of $1/5^{rm th}$ that used in the standard
    resolution runs (solid curves) used 
    in this paper. There are differences in the central stellar density -- within the galaxy,
    such that there is a higher density in the lower resolution run -- and in the gas
    density at intermediate radii, but the shape and amplitude of the stellar density
    within the ESC is similar between the different resolutions.}
  \label{fig:density_profile_cdm_z0}
\end{figure}


\begin{thebibliography}{0}

\bibitem[Abadi, Navarro, \& Steinmetz(2006)]{abadi.etal.2006} Abadi M.~G., Navarro J.~F., Steinmetz M., 2006, MNRAS, 365, 747
\bibitem[Amorisco(2015)]{amorisco.2015} Amorisco, N.~C.\ 2015, arXiv:1511.08806
\bibitem[Anderhalden et al.(2013)]{anderhalden.etal.2013} Anderhalden, D., Schneider, A., Macci{\`o}, A.~V., Diemand, J., \& Bertone, G.\ 2013, JCAP, 3, 014 
\bibitem[Angulo et al.(2009)]{angulo.etal.2009} Angulo R.~E., Lacey C.~G., Baugh C.~M., Frenk C.~S., 2009, MNRAS, 399, 983 
\bibitem[Angulo, Hahn, \& Abel(2013)]{angulo.etal.2013} Angulo R.~E., Hahn O., Abel T., 2013, MNRAS, 434, 3337 
\bibitem[Benson(2005)]{benson.2005} Benson A.~J., 2005, MNRAS, 358, 551 
\bibitem[Benson et al.(2013)]{benson.2013} Benson, A.~J., Farahi, A., Cole, S., et al.\ 2013, MNRAS, 428, 1774 
\bibitem[Binney \& Tremaine(2008)]{binney.tremaine.2008} Binney, J., \& Tremaine, S.\ 2008, Galactic Dynamics: Second Edition, by James Binney and Scott Tremaine.~ISBN 978-0-691-13026-2 (HB).~Published by Princeton University Press, Princeton, NJ USA, 2008.,  
\bibitem[Bode et al.(2001)]{bode.etal.2001} Bode, P., Ostriker, J.~P., \& Turok, N.\ 2001, ApJ, 556, 93 
\bibitem[Boley et al.(2009)]{boley.etal.2009} Boley, A.~C., Lake, G., Read, J., \& Teyssier, R.\ 2009, ApJL, 706, L192 
\bibitem[Boylan-Kolchin et al.(2011)]{boylan-kolchin.etal.2011} Boylan-Kolchin, M., Bullock, J.~S., \& Kaplinghat, M.\ 2011, MNRAS, 415, L40
\bibitem[Brooks \& Zolotov(2014)]{brooks.zolotov.2014} Brooks, A.~M., \& Zolotov, A.\ 2014, \apj, 786, 87
\bibitem[Bullock \& Johnston(2005)]{bullock.johnston.2005} Bullock, J.~S., \& Johnston, K.~V.\ 2005, ApJ, 635, 931 
\bibitem[Carlin et al.(2016)]{carlin.etal.2016} Carlin J.~L., Beaton R.~L., Mart{\'{\i}}nez-Delgado D., Gabany R.~J., 2016, ASSL, 420, 219 
\bibitem[Col{\'{\i}}n et al.(2008)]{colin.etal.2008} Col{\'{\i}}n, P., Valenzuela, O., \& Avila-Reese, V.\ 2008, ApJ, 673, 203 
\bibitem[Col{\'{\i}}n et al.(2015)]{colin.etal.2015} Col{\'{\i}}n, P., Avila-Reese, V., Gonz{\'a}lez-Samaniego, A., \& Vel{\'a}zquez, H.\ 2015, ApJ, 803, 28 
\bibitem[Cooper et al.(2010)]{cooper.etal.2010} Cooper, A.~P., Cole, S., Frenk, C.~S., et al.\ 2010, MNRAS, 406, 744 
\bibitem[Cooper et al.(2013)]{cooper.etal.2013} Cooper A.~P., D'Souza R., Kauffmann G., Wang J., Boylan-Kolchin M., Guo Q., Frenk C.~S., White S.~D.~M., 2013, MNRAS, 434, 3348
\bibitem[Crnojevi{\'c} et al.(2015)]{crnojevic.etal.2015} Crnojevi{\'c}, D., Sand, D.~J., Spekkens, K., et al.\ 2015, arXiv:1512.05366 
\bibitem[Dehnen \& Aly(2012)]{dehnen.aly.2012} Dehnen W., Aly H., 2012, MNRAS, 425, 1068 
\bibitem[Diemand, Kuhlen, \& Madau(2007)]{diemand.etal.2007} Diemand J., Kuhlen M., Madau P., 2007, ApJ, 667, 859 
\bibitem[Dutton et al.(2016)]{dutton.etal.2016} Dutton, A.~A., Macci{\`o}, A.~V., Frings, J., et al.\ 2016, \mnras, 457, L74
\bibitem[Efstathiou et al.(1985)]{efstathiou.etal.1985} Efstathiou, G., Davis, M., White, S.~D.~M., \& Frenk, C.~S.\ 1985, ApJS, 57, 241
\bibitem[Elbert et al.(2015)]{elbert.etal.2015} Elbert, O.~D., Bullock, J.~S., Garrison-Kimmel, S., et al.\ 2015, \mnras, 453, 29
\bibitem[Font et al.(2011)]{font.etal.2011} Font, A.~S., McCarthy, I.~G., Crain, R.~A., et al.\ 2011, MNRAS, 416, 2802 
\bibitem[Freeman \& Bland-Hawthorn(2002)]{freeman.bland-hawthorn.2002} Freeman, K., \& Bland-Hawthorn, J.\ 2002, ARA\&A, 40, 487 
\bibitem[Fry et al.(2015)]{fry.etal.2015} Fry A.~B., et al., 2015, MNRAS, 452, 1468 
\bibitem[Gao et al.(2004)]{gao.etal.2004} Gao L., White S.~D.~M., Jenkins A., Stoehr F., Springel V., 2004, MNRAS, 355, 819 
\bibitem[Garrison-Kimmel et al.(2014)]{garrison-kimmel.etal.2014} Garrison-Kimmel, S., Boylan-Kolchin, M., Bullock, J.~S., \& Lee, K.\ 2014, MNRAS, 438, 2578 
\bibitem[Garrison-Kimmel et al.(2016)]{garrison-kimmel.etal.2016} Garrison-Kimmel, S., Bullock, J.~S., Boylan-Kolchin, M., \& Bardwell, E.\ 2016, arXiv:1603.04855 
\bibitem[Gilbert et al.(2012)]{gilbert.etal.2012} Gilbert K.~M., et al., 2012, ApJ, 760, 76 
\bibitem[Governato et al.(2010)]{governato.etal.2010} Governato, F., Brook, C., Mayer, L., et al.\ 2010, Nature, 463, 203 
\bibitem[Governato et al.(2015)]{governato.etal.2015} Governato F., et al., 2015, MNRAS, 448, 792 
\bibitem[Helmi \& White(1999)]{helmi.white.1999} Helmi, A., \& White, S.~D.~M.\ 1999, MNRAS, 307, 495 
\bibitem[Helmi(2008)]{helmi.2008} Helmi, A.\ 2008, A\&ARv, 15, 145 
\bibitem[Herpich et al.(2014)]{herpich.etal.2014} Herpich, J., Stinson, G.~S., Macci{\`o}, A.~V., et al.\ 2014, MNRAS, 437, 293 
\bibitem[Hobbs et al.(2013)]{hobbs.etal.2013} Hobbs, A., Read, J., Power, C., \& Cole, D.\ 2013, MNRAS, 434, 1849 
\bibitem[Hoffman et al.(2012)]{hoffman.etal.2012} Hoffman, Y., Metuki, O., Yepes, G., et al.\ 2012, MNRAS, 425, 2049 
\bibitem[Ishiyama et al.(2013)]{ishiyama.etal.2013} Ishiyama, T., Rieder, S., Makino, J., et al.\ 2013, ApJ, 767, 146
\bibitem[Jiang et al.(2015)]{jiang.etal.2015} Jiang L., Cole S., Sawala T., Frenk C.~S., 2015, MNRAS, 448, 1674 
\bibitem[Johnston et al.(2008)]{johnston.etal.2008} Johnston, K.~V., Bullock, J.~S., Sharma, S., et al.\ 2008, ApJ, 689, 936-957 
\bibitem[Katz et al.(1996)]{katz.etal.1996} Katz, N., Weinberg, D.~H., \& Hernquist, L.\ 1996, ApJS, 105, 19 
\bibitem[Kennicutt(1998)]{kennicutt.1998} Kennicutt, R.~C., Jr.\ 1998, ApJ, 498, 541 
\bibitem[Khochfar \& Burkert(2006)]{khochfar.burkert.2006} Khochfar S., Burkert A., 2006, A\&A, 445, 403 
\bibitem[Klypin et al.(1999)]{klypin.etal.1999b} Klypin A., Kravtsov A.~V., Valenzuela O., Prada F., 1999, ApJ, 522, 82 
\bibitem[Knebe et al.(2008)]{knebe.etal.2008} Knebe, A., Arnold, B., Power, C., \& Gibson, B.~K.\ 2008, MNRAS, 386, 1029 
\bibitem[Komatsu et al.(2011)]{komatsu.etal.2011} Komatsu, E., Smith, K.~M., Dunkley, J., et al.\ 2011, ApJS, 192, 18 
\bibitem[Lada \& Lada(2003)]{lada.lada.2003} Lada, C.~J., \& Lada, E.~A.\ 2003, \araa, 41, 57 
\bibitem[Lewis et al.(2000)]{lewis.etal.2000} Lewis, A., Challinor, A., \& Lasenby, A.\ 2000, ApJ, 538, 473 
\bibitem[LSST Science Collaboration et al.(2009)]{lsst} LSST Science Collaboration, Abell, P.~A., Allison, J., et al.\ 2009, arXiv:0912.0201 
\bibitem[Mackey et al.(2010)]{mackey.etal.2010} Mackey A.~D., et al., 2010, ApJ, 717, L11 
\bibitem[Mart{\'{\i}}nez-Delgado et al.(2010)]{martinez-delgado.etal.2010} Mart{\'{\i}}nez-Delgado D., et al., 2010, AJ, 140, 962 
\bibitem[Mashchenko et al.(2008)]{mashchenko.etal.2008} Mashchenko, S., Wadsley, J., \& Couchman, H.~M.~P.\ 2008, Science, 319, 174 
\bibitem[McCarthy et al.(2012)]{mccarthy.etal.2012} McCarthy, I.~G., Font, A.~S., Crain, R.~A., et al.\ 2012, MNRAS, 420, 2245 
\bibitem[McConnachie et al.(2009)]{mcconnachie.etal.2009} McConnachie A.~W., et al., 2009, Nature, 461, 66 
\bibitem[Monachesi et al.(2013)]{monachesi.etal.2013} Monachesi A., et al., 2013, ApJ, 766, 106 
\bibitem[Moore et al.(1999)]{moore.etal.1999} Moore B., Ghigna S., Governato F., Lake G., Quinn T., Stadel J., Tozzi P., 1999, ApJ, 524, L19 
\bibitem[Pacucci et al.(2013)]{pacucci.2013} Pacucci, F., Mesinger, A., \& Haiman, Z.\ 2013, MNRAS, 435, L53 
\bibitem[Power et al.(2003)]{power.etal.2003} Power, C., Navarro, J.~F., Jenkins, A., et al.\ 2003, MNRAS, 338, 14
\bibitem[Power(2013)]{power.2013} Power, C.\ 2013, PASA, 30, 53 
\bibitem[Power et al.(2014a)]{power.2014} Power, C., Wynn, G.~A., Robotham, A.~S.~G., Lewis, G.~F., \& Wilkinson, M.~I.\ 2014a, arXiv:1406.7097 
\bibitem[Power et al.(2014b)]{power.etal.2014} Power, C., Read, J.~I., \& Hobbs, A.\ 2014b, MNRAS, 440, 3243 
\bibitem[Radburn-Smith et al.(2011)]{radburn-smith.etal.2011} Radburn-Smith D.~J., et al., 2011, ApJS, 195, 18 
\bibitem[Rashkov et al.(2013)]{rashkov.etal.2013} Rashkov, V., Pillepich, A., Deason, A.~J., et al.\ 2013, \apjl, 773, L32
\bibitem[Read et al.(2006)]{read.etal.2006} Read, J.~I., Pontzen, A.~P., \& Viel, M.\ 2006, MNRAS, 371, 885 
\bibitem[Read \& Hayfield(2012)]{read.hayfield.2012} Read J.~I., Hayfield T., 2012, MNRAS, 422, 3037 
\bibitem[Reed et al.(2005)]{reed.etal.2005} Reed D., Governato F., Quinn T., Gardner J., Stadel J., Lake G., 2005, MNRAS, 359, 1537
\bibitem[Rodriguez-Gomez et al.(2016)]{rodriguez-gomez.etal.2016} Rodriguez-Gomez, V., Pillepich, A., Sales, L.~V., et al.\ 2016, \mnras, 458, 2371
\bibitem[Rodriguez-Puebla et al.(2016)]{rodriguez-puebla.etal.2016} Rodriguez-Puebla, A., Behroozi, P., Primack, J., et al.\ 2016, arXiv:1602.04813 
\bibitem[Salpeter(1955)]{salpeter.1955} Salpeter, E.~E.\ 1955, ApJ, 121, 161 
\bibitem[Sawala et al.(2016a)]{sawala.etal.2016a} Sawala, T., Frenk, C.~S., Fattahi, A., et al.\ 2016, MNRAS, 457, 1931 
\bibitem[Sawala et al.(2016b)]{sawala.etal.2016b} Sawala, T., Frenk, C.~S., Fattahi, A., et al.\ 2016, MNRAS, 456, 85 
\bibitem[Scannapieco et al.(2011)]{scannapieco.etal.2011} Scannapieco C., White S.~D.~M., Springel V., Tissera P.~B., 2011, MNRAS, 417, 154 
\bibitem[Schmidt(1959)]{schmidt.1959} Schmidt, M.\ 1959, ApJ, 129, 243 
\bibitem[Schneider et al.(2013)]{schneider.2013} Schneider, A., Smith, R.~E., \& Reed, D.\ 2013, MNRAS, 433, 1573 
\bibitem[Schneider et al.(2014)]{schneider.etal.2014} Schneider, A., Anderhalden, D., Macci{\`o}, A.~V., \& Diemand, J.\ 2014, MNRAS, 441, L6 
\bibitem[Searle \& Zinn(1978)]{searle.zinn} Searle L., Zinn R., 1978, ApJ, 225, 357 
\bibitem[Sembolini et al.(2016)]{sembolini.etal.2016} Sembolini, F., Yepes, G., Pearce, F.~R., et al.\ 2016, MNRAS,  
\bibitem[Smith \& Markovic(2011)]{smith.markovic.2011} Smith, R.~E., \& Markovic, K.\ 2011, PhRvD, 84, 063507 
\bibitem[Springel(2005)]{springel.2005} Springel, V.\ 2005, MNRAS, 364, 1105
\bibitem[Springel et al.(2008)]{springel.etal.2008} Springel, V., Wang, J., Vogelsberger, M., et al.\ 2008, MNRAS, 391, 1685
\bibitem[Tormen, Diaferio, \& Syer(1998)]{tormen.etal.1998} Tormen G., Diaferio A., Syer D., 1998, MNRAS, 299, 728 
\bibitem[Trujillo \& Bakos(2013)]{trujillo.etal.2013} Trujillo I., Bakos J., 2013, MNRAS, 431, 1121 
\bibitem[van Dokkum et al.(2014)]{dragonfly} van Dokkum, P.~G., Abraham, R., \& Merritt, A.\ 2014, ApJL, 782, L24 
\bibitem[Wetzel(2011)]{wetzel.2011} Wetzel A.~R., 2011, MNRAS, 412, 49 
\bibitem[Wetzel et al.(2016)]{wetzel.etal.2016} Wetzel, A.~R., Hopkins, P.~F., Kim, J.-h., et al.\ 2016, arXiv:1602.05957
\bibitem[Xie \& Gao(2015)]{xie.gao.2015} Xie, L., \& Gao, L.\ 2015, MNRAS, 454, 1697
\bibitem[Zavala et al.(2013)]{zavala.etal.2013} Zavala, J., Vogelsberger, M., \& Walker, M.~G.\ 2013, \mnras, 431, L20
\bibitem[Zel'dovich(1970)]{zeldovich.1970} Zel'dovich Y.~B., 1970, A\&A, 5, 84 
\bibitem[Zhu et al.(2016)]{zhu.etal.2016} Zhu, Q., Marinacci, F., Maji, M., et al.\ 2016, \mnras, 458, 1559
\bibitem[Zolotov et al.(2009)]{zolotov.etal.2009} Zolotov A., Willman B., Brooks A.~M., Governato F., Brook C.~B., Hogg D.~W., Quinn T., Stinson G., 2009, ApJ, 702, 1058 
  
\end{thebibliography}
\end{document}